\documentstyle[12pt]{article}
\newlength{\dinwidth}
\newlength{\dinmargin}
 \def\bbbr{{\rm I\!R}}
 \def\bbbone{{\mathchoice {\rm 1\mskip-4mu l}
 {\rm 1\mskip-4mu l}
 {\rm 1\mskip-4.5mu l} {\rm 1\mskip-5mu l}}}
 \def\bbbz{{\mathchoice {\hbox{$\sf\textstyle Z\kern-0.4em Z$}}
 {\hbox{$\sf\textstyle Z\kern-0.4em Z$}}
 {\hbox{$\sf\scriptstyle Z\kern-0.3em Z$}}
 {\hbox{$\sf\scriptscriptstyle Z\kern-0.2em Z$}}}}
\def\be{\begin{equation}}
\def\ba{\begin{eqnarray}}
\def\ee{\end{equation}}
\def\ea{\end{eqnarray}}
\def\nn{\nonumber}
\def\A{{\cal A}}             % interpolation kernel
\def\D{{\cal D}}             % functional measure
\def\H{{\cal H}}             % good old Hamilton
\def\N{{\cal N}}             % normalization factor
\def\V{{\cal V}}             % multiplication operator in section 4
\def\vp{\varphi }            % fundamental fields
\def\tr{{\rm Tr}}            % trace
\def\Z{{\tt z}}              % factor \ref{z} in section 1
\def\hat#1{\widehat{#1}}
\def\tilde#1{\widetilde{#1}}
\def\L{\Lambda}                       % lattices
\def\A{{\cal A}}                      % interpolation kernels
\def\Dirac{\mbox{$\not\!\nabla$}}     % slashed Dirac operator
\def\G{\Gamma}                        % high frequency propagator
\def\Lap{\Delta}                      % Laplace operator
\def\UB{{\cal U}}                     % block lattice gauge field
\def\a{\alpha}                        % spinor index
\def\CG{{\cal G}}                     % staggered fermion symmetry group
\def\bi{\begin{itemize}}     % itemization
\def\ei{\end{itemize}}
\def\bn{\begin{enumerate}}   % enumeration
\def\en{\end{enumerate}}
\def\bdm{\begin{eqnarray*}}  % displaymath
\def\edm{\end{eqnarray*}}
\def\es{\enspace}            % standard space at end of eq.

\def\eq#1{{(\ref{#1})}}      % brackets for equation references
\def\wh#1{\widehat{#1}}
\def\p{\prime}
\def\eh{\frac{1}{2}}         % 1/2 (in equations only)

\def\me{{-1}}

\newcommand{\skipp}[1]{\mbox{\hspace{#1 ex}}}
\def\Tr{{\rm Tr}}                      % trace
\def\Arno{|A_{r n o}|^{2}}
\def\Gnp{\Gamma_{n \phi_{r}}(\Phi)}
\def\Gmp{\Gamma_{m \phi_{r}}(\Phi)}
\def\Arzmu{A_{r z \mu}}
\def\Kn{\hat{k}^{2}_{n}}
\def\Arz{A_{r z - \mu \mu}}

\newcommand{\ds}{\displaystyle \sum}   % sum symbol
 \newcommand{\subs}[1]{\mbox{\protect\scriptsize\rm #1}}
 \newcommand{\mcr}{m_{\subs{cr}}^2}
 \newcommand{\Cstar}{C^{\ast}}
 \newcommand{\Dm}{\triangle m^2}
 \newcommand{\Lone}{\Lambda^1}
 \newcommand{\Lnull}{\Lambda^0}
 \newcommand{\Ltwo}{\Lambda^2}
\begin{document}
 \noindent{\tt DESY $92-070$ \hfill ISSN $0418-9833$}\\
 {\tt May 1992}

 \renewcommand{\thefootnote}{{\protect\fnsymbol{footnote}}}

 \begin{center}
 \mbox{}
 \vspace{3cm}

 {\LARGE  Effective Field Theories} \\

 \vspace{1.5cm}
             G. Mack,
             T. Kalkreuter,
             G. Palma\footnote{On leave of absence from Universidad de
             Santiago de Chile, Casilla 307, Correo 2, Santiago-Chile},
             and M. Speh \\  \smallskip
            II. Institut f\"ur Theoretische Physik,
            Universit\"at Hamburg,\\
            Luruper Chaussee 149, 2000 Hamburg 50
 \vspace{1.5cm}
 \mbox{}

 \end{center}

 \vfill

\begin{abstract}
Effective field theories encode the predictions of a quantum field
theory at low energy. The effective theory has a fairly low
ultraviolet cutoff. As a result, loop corrections are small, at least
if the effective action contains a term which is quadratic in the
fields, and physical predictions can be read straight from the effective
Lagrangean.

Methods will be discussed how to compute an effective low energy
action from a given fundamental action, either analytically or
numerically, or by a combination of both methods. Basically,
the idea is to integrate out the high frequency components of fields.
This requires the choice of a ``blockspin'', i.e. the specification
of a low frequency field as a function of the fundamental fields.
These blockspins will be the fields of the effective field theory.
The blockspin need not be a field of the same type as one of the
fundamental fields, and it may be composite. Special features of
blockspins in nonabelian gauge theories will be discussed in some
detail.

In analytical work and in multigrid updating schemes
one needs interpolation kernels $\A$ from coarse to fine grid in
addition to the averaging kernels $C$ which determines the blockspin.
A neural net strategy for finding optimal kernels is presented.

Numerical methods are applicable to obtain actions of
effective theories on lattices
of finite volume. The  special case of a ``lattice'' with a single site
(the constraint effective potential)
is of particular interest. In a Higgs model, the effective action
reduces in this case to the free energy, considered as  a function of
a gauge covariant magnetization. Its shape determines the phase
structure of the theory. Its loop expansion with and without gauge
fields can be used to determine finite size corrections to numerical
data.
\end{abstract}

 \vfill
 \mbox{}

 \renewcommand{\thefootnote}{\arabic{footnote}}
 \setcounter{footnote}{0}

  \newpage

\tableofcontents
 \mbox{}\vfill
\section{What are Effective Field Theories?
         \label{SECTION_WHAT}}
We work in the context of Euclidean field theory.
\begin{itemize}
\item An effective field theory encodes the predictions of a quantum
      field theory at \it low energies \rm .
\item It is a \it field theory \rm with a \it  low UV-cutoff \rm
      $ a_0^{-1}$ - of the order of the relevant particle masses. For
      instance it may live on a \it lattice \rm of lattice spacing
      $a_0$.
\item Ideally one should be able to read the physics straight from
      the \it effective action \rm (= action of the effective theory)
      because loop corrections to tree amplitudes are small at
      sufficiently low UV-cutoff.
\end{itemize}

\newpage
\subsubsection*{Problem to be discussed}
      Start from a given \it fundamental action \rm (on the continuum
      or on a lattice of small lattice spacing $a \ll a_0 $)
\begin{itemize}
\item How does one {\it define and compute the effective action}?
\end{itemize}
 Sometimes effective actions are guessed on the basis of known
 properties.

 {\bf Example:} chiral theories (nonlinear $\sigma $-models) as
 effective theories of mesons. They embody the low energy theorems
 implied by PCAC.\footnote{ PCAC = partially
 conserved axial vector current.}
 These low energy theorems
 are sufficiently restrictive to fix the
  effective action at really low energy (i.e. low cutoff) up to a few
 coupling constants. By the definition of an effective theory, the
 question of the removal of the UV-cutoff does not pose itself.
 Nonrenormalizable $\sigma $-models are therefore acceptable.

 But ideally, one would like not to guess the effective theory, but to
 justify it from the given fundamental theory.
 \subsection*{Definition of the effective action in terms of the
  fundamental theory
  \label{SUBSECTION_WHAT_DEFINITION}}
 The effective action will be a function of \it low frequency fields
 $ \Phi $ \rm - for instance, fields $\Phi $ on a lattice of lattice
 spacing $a_0$. They should be defined as {\it functions of the
 fundamental fields} $\vp $,
 $$ \Phi = C(\vp ) \ , $$
 If $C$ is linear, we write instead
 $$ \Phi = C \vp   \ . $$
 $C$ will be called the averaging operator
 and $\Phi$ is called {\it blockspin}.
 Later on we will consider
                  fundamental theories which are gauge theories, and we
 will             distinguish in notation between matter fields $\vp $
 and gauge fields $U$. In general the low frequency matter fields
 will be functions not only of the matter fields $\vp $
 but will also depend on the gauge fields,
 $$ \Phi = C(U,\vp ) \ \ , \mbox{ or }  \ \
    \Phi = C(U)\vp  \ . $$
 The definition of the effective action $\H_{eff} $ in terms of the
 fundamental Euclidean action $\H (\vp  )$ reads
 \footnote{Euclidean field theory may be regarded as a classical
 statistical mechanics. In this language, the action becomes the
 Hamiltonian. Therefore we use the letter $\H $.}
\be  e^{-\H_{eff}(\Phi)}= \int \D \vp
  \  \delta (\Phi - C(\vp) )\ e^{-\H (\vp)} \label{DefHeff}
\ee
If the effective theory lives on a lattice $\Lambda_0$, then
$$ \delta (\Phi -C(\vp ))= \prod_{\alpha }\prod_{x\in \Lambda_0}
  \delta (\Phi_{\alpha }(x)-C_{\alpha }(\vp )(x)) $$
when $\Phi $ has components  $\Phi_{\alpha }$.
The 1-dimensional $\delta $-function admits the integral representation
$$ \delta (\xi ) =
                   \frac 1{2\pi }\int_{-\infty }^{\infty }
   dk\ e^{ik\xi }\ . $$
This formula, or (better) an approximation of the $\delta$-function by
a Gaussian
$$ \delta_{\beta }(\xi ) = \left( \frac{2\pi }{\beta } \right)
        e^{-\xi^2/2\beta } =
                   \frac 1{2\pi }\int_{-\infty }^{\infty }
   dk\ e^{ik\xi }\ e^{-\beta k^2 /2} \quad (\beta\ \
       \mbox{small})  $$
 can also be used when there are \it Fermi fields. \rm
\subsubsection*{Problems}
 \begin{itemize}
\item On what \it number and kinds of fields \rm $\Phi_{\alpha }$ should
 $\H_{eff} $ depend ?
\item How to choose the \it averaging operator \rm $C$.\\
      Averaging over space time regions will be involved in getting
      low frequency fields.
\item How to {\it compute the functional integral}~\eq{DefHeff}
      by analytical or numerical means.
\item  How to \it detect a bad choice of $C$ or $\Phi $.\rm
\end{itemize}
The answer to the last question is given by the requirement that \it
the effective action must have good locality properties. \rm

Let us emphasize that  $\H_{eff} $ is {\it exact} in the sense that it
yields exact expectation values for observables $O$ which depend on the
fundamental fields $\vp $ only through
$\Phi_{\alpha } = C_{\alpha }(\vp ) $ so that $O = F(C(\vp ))$.
\ba \langle O \rangle
        &=& \frac 1Z\int \D\vp\ F(C(\vp ))\ e^{-\H(\vp )} \nn\\
&=& \frac 1Z\int \D\Phi\  F(\Phi )\ e^{-\H_{eff}(\Phi )}\es .
\label{O}
\ea
We illustrate the issues involved as well as the dangers of a bad choice
of $\Phi $ or $C$ at an example which is well understood.

{\bf Example: } Discrete Gaussian model.

We consider first the case where the fundamental theory is a discrete
Gaussian model in 3 dimensions. It lives on a lattice $\Lambda $
of lattice spacing  $a$.  This model is the dual transform of a
$U(1)$ lattice gauge theory in 3 dimensions. An exact computation
of its effective action by convergent series expansions was used in
\cite{WHAT_GM} to give a rigorous proof that the 3-dimensional
$U(1)$-lattice gauge theory shows linear confinement for all values of
the gauge coupling constant $g$,with $ \beta = 4\pi^2/g^2$.
\footnote{ The convergent expansions are iterated Mayer expansions.
They combine expansions at various length scales. Similar expansions
can be written down which live on a multigrid \cite{MPordt}
                                              - cp. later.}

The points of the fundamental lattice will be denoted by letters $z$.
The fundamental field is \it integer valued \rm ,
$$ n(z) \in \bbbz \ , $$
and the fundamental action is
\footnote{We use lattice notations as follows on a $d$-dimensional
                                                   lattice $\Lambda $
of lattice spacing $a$:
$ \int_z = a^d \sum_{z\in \Lambda } $ and
$ \nabla_{\mu } f (z) = a^{-1} \left[ f (z+e_{\mu })- f(z)\right ]$,
$e_{\mu }$= lattice vector of length $a$ in $\mu $-direction.
}
$$ \H (n) = \frac 1{2\beta }\int_z [\nabla_{\mu }n(z)]^2 \ . $$
The difficult case is when $\beta /a $ \it is large \rm (i.e. weak gauge
coupling $g$.) Note that $\beta /a $ is dimensionless in 3 dimensions.

A choice was made to use a Pauli-Villars cutoff for the effective theory,
so that the effective theory lives on the original lattice $\Lambda $,
to begin with. The cutoff $a_0^{-1}$ was chosen to be of the order of the
ultimate physical mass $m$ of the theory, in agreement with the
philosophy of effective field theory. But it is essential to choose
as the low frequency field a \it real field \rm
$$ \Phi (z) \ \mbox{real } \ . $$
This illustrates the point that \it the fields $\Phi $ of the
effective theory need not be fields of the same type as those
in the fundamental theory \rm .

One finds that a local sine-Gordon action is an excellent approximation
to the effective action when $\beta $ is large
\be \H_{eff }(\Phi ) = \int_z\left( \frac 1{2\beta }
  \left[ \nabla_\mu \Phi (z) \right]^2 +
    2\Z \left[ 1-\cos \Phi (z) \right] \right)
\ee
with
\ba \Z &=& a^{-3}\exp\left[-\beta v_{Cb}(0)\right]
\label{z}
\quad \mbox{for large } \beta \ , \\
 v_{Cb}(0) &=& 0.2527\  a^{-1}\nn \  . \ea
$v_{Cb}(0)$ is the Coulomb potential at zero distance on the
3-dimensional lattice.
(We chose not to write the terms in $\H_{eff} $
         which implement  the Pauli-Villars cutoff.)

We proceed to a discussion of the physics of the model on the basis of
of the effective action.

1) {\bf Mass: } The $[1-\cos]$ potential has its minima at
$\Phi = 2\pi \cdot$integer. We shall see in a moment that
tunneling between different minima is very much suppressed, so that
the global symmetry
$$
\Phi (x) \mapsto \Phi (x) + 2\pi N\ ,  \ \ N \mbox{ integer }
$$
is spontaneously broken, and we may restrict attention to one of the
minima, e.g. $\Phi = 0 $, and expand around it. Set
\ba
     \Phi &=& \beta^{1/2}\ \Psi \quad ,\nn \\
     m^2&=& \Z\ 2\beta =
                 2\beta\ a^{-3}\exp\left[-\beta v_{Cb}(0)\right]
\es .\label{m} \ea
Then the effective action takes the form
\be \H_{eff }= \frac 12\int_z \left(
  \left[ \nabla_\mu \Psi (z) \right]^2 +
    m^2\ \frac 12 \ \Psi (z)^2 + \left[\cdots\right]  \right)  \es . \nn \ee
The nonlinear terms $\left[\cdots\right]$ are small for large $\beta $
 (provided
 $\Psi $ is sufficiently near to the selected minimum so that the
effective action is sufficiently small to give $\Psi $ a
nonnegligible probability). The effective theory is therefore very
nearly a free field theory with a nonvanishing mass $m$ which
is given by~\eq{m}. Numerical simulations confirm this
\cite{WHAT_KM}.

The low lying excitations of the model are therefore spin waves
with a finite mass which is exponentially small
in units of the original inverse lattice spacing $a^{-1}$.
This is easy to read off the effective action, but is not at all
apparent from the fundamental action.

2) {\bf Tunneling: } Because of the UV-cutoff we can qualitatively think
of the theory as living on a lattice $\Lambda_0$ of lattice spacing
$a_0= \mbox{ const. }\cdot m^\me$. The effective action becomes
\ba   \H_{eff } &=&\sum_{x\in \Lambda_0} \frac {a_0^3}{2\beta }
  \left[ \nabla_\mu \Phi (x) \right]^2 +
   2\Z_0 \left[ 1-\cos \Phi (x) \right]  \label{EffsinG}  \\
\Z_0 &=& a_0^3\Z\ =\ \mbox{ const. }^\p\  m^{-3}\Z \\
           &=&\mbox{ const } ^{\p\p} \beta^{-3/2}
        \exp\left[\beta v_{Cb}(0)/2\right] \ea
We see that ${\Z_0 } \mapsto \infty $ when $\beta \mapsto \infty$.
Therefore the hills of the $2\Z_0\left[ 1-\cos \Phi \right]$~-~potential
become very high when $\beta $ becomes large. As a result, tunneling
becomes very much suppressed.

The height and widths of the hills of the potential determine the
surface tension of the discrete Gaussian model (the free energy
per area of an interphase between domains where
$\langle \Phi \rangle = 2\pi N $
with different $N$). This surface tension is equal to the string tension
$\alpha $ of
the dual $U(1)$-gauge theory. Treating the tunneling in  a semiclassical
approximation results in
\be \alpha = 8 m\beta^{-1} \ . \ee
Note that
$$ \alpha /m^2 \mapsto \infty \ \ \mbox{as } \beta \mapsto \infty \ . $$
Thus \it there are two different physical mass scales in this theory.\rm

3) {\bf 2-dimensional discrete Gaussian model:}

Next we turn to the  2-{\it dimensional discrete Gaussian model}.
This model is the dual transform of the XY-model (plane rotator) in
2 dimensions with Villain action.

A rigorous computation of the effective action of this model has not
been reported in the literature, but the method used for the
3-dimensional model can be expected to carry over for large $\beta $.
In 2 dimensions, $\beta $ is dimensionless.

The effective action comes out the same as in 3 dimension, except for
the appearance of different powers of $a$ (for dimensional reasons) and
the following difference in the expression for $\Z $.
Expression~\eq{z} results from an approximation which is
legitimate in 3 dimensions, but not in 2. Originally there stands
the Yukawa potential $v_{\mu }(0)$ at zero distance in place of
$ v_{Cb}(0)$, with mass $\mu $ equal to the Pauli-Villars cutoff
$a_0^{-1}$. We are interested in the limit where $\mu a$ is very small.
In 3 dimensions, the zero mass limit exists, and we may approximate
$v_{\mu }(0)$ by $v_{Cb}(0)$. But in 2 dimensions, the zero mass limit
does not exist. Instead,
$ v_{\mu }(0) $ increases logarithmically with $(\mu a)^{-1}$ as the
cutoff $\mu = a_0^{-1}$ is lowered. As a result,
\be
 {\Z_0} = a_0^2\ \Z =(a\mu )^{-2} \exp\left[-\beta v_{\mu }(0)/2\right]
\ee
decreases when the cutoff is lowered, if $\beta $ is large enough,
and the mass in tree approximation decreases faster than the cutoff
$\mu $. Therefore, the cutoff can be lowered indefinitely.
Since $\Z_0$ decreases indefinitely, the
effective action tends to a massless free field theory in the limit
of very low cutoff, provided $\beta $ is large enough. This is the
Kosterlitz-Thouless phase of the model.

There is no particular reason to believe that the formula~\eq{EffsinG}
for the effective action remains accurate for intermediate values
of $\beta $. But the nature of the Kosterlitz-Thouless phase
phase transition could be read off, if one had an accurate expression
for the effective action at intermediate $\beta $.

\subsubsection*{Lessons from the model}
\begin{itemize}
\item The physics can readily be read from the effective action, but is
not visible in the fundamental action.
\item A good choice of the blockspin $\Phi $ (real as opposed to
integer) is crucial.
\end{itemize}

If we had chosen an integer blockspin, the spin waves of zero or
small mass m                                           as the dominant
low energy excitations would not be represented by fields in the action.
Moreover, one would meet a disaster while calculating the effective
action. If one tried to compute it by lowering the cutoff step by step
(which is what an iterated Mayer expansion does), the action would
become nonlocal over distances $m^{-1}$ while the UV-cutoff $\mu $
is still very much higher than $m$.  A good choice of the blockspin
is
$$ \Phi (z) \approx \mbox{average of }n(w)\mbox{ over a cube of side }
    a_0 $$
using a smoothened step function for the cube.
\subsection*{Stages of complexity of the effective action
            \label{SUBSECTION_WHAT_STAGES}}
To be specific,
let us start from the fundamental action of a Higgs model with
nonabelian gauge fields living on $\Lambda$. $\Lambda $ may be either
the continuum or a lattice of small lattice spacing.
 We seek an effective theory on a lattice
$\Lambda_0$ of lattice spacing $a_0$.

We identify the points $x\in \Lambda_0$ with hypercubes  in the
fundamental lattice $\Lambda$.

\it These hypercubes are called  blocks\rm.
\rm Sometimes we will also regard $\Lambda_0$ as a sublattice of
$\Lambda$, so that $x\in \Lambda_0$ fixes a site in $\Lambda $, e.g.
the block center.

One may envisage different kinds of and approximations to the effective
action which require different kinds of blockspins

\begin{enumerate}
\item Ignore both the gauge field $U$ and the fermion fields $\psi $.\\
Approximate the fundamental theory by a $\lambda\vp^4$-theory.
Choose the block average of the scalar field as blockspin \cite{WHAT_GK}
$$ \Phi_{\alpha } (x) = \int_z C(x,z)\ \vp_{\alpha }(z) =
Z_0^{-1/2}(a_0)\  av_{z \in x} \vp_{\alpha }(z) \ . $$
with $C (x,z) \propto $ step function. A cutoff dependent
factor $Z_0^{1/2}(a_0)$ (wave function renormalization)
has been included with the aim to normalize the kinetic term in the
effective action in the standard way.
\item Ignore the gauge field, but consider fermions. \\
      An effective action which has chiral symmetry
      requires $2^{d/2}$ flavours $f$ in $d$ dimensions (for even $d$).
      Otherwise the effective action will inevitably become nonlocal
      (Nielsen-Ninomiya theorem \cite{FERMIONS_NIELSEN_NINOMIYA}).
      Given
      fundamental fermion fields with  $2^{d/2}$ flavours, we may block
      to staggered fermions on $\Lambda_0$ of lattice spacing
      $\frac{1}{2}a_0$.                     $\Lambda_0$ decomposes
      into elementary cells of sidelength $a_0$ containing $2^d$ sites
      each. Only translations by integer multiples of $a_0$ are true
      translations. Sites within one elementary cell carry different
      pseudoflavour (a combination of spin and flavour). A suitable
      blockspin is of the form
\be
\Psi (x) = \int_z \sum_{f,a} C_a^f(x,z)\ \psi^f_a(z)
\es
\label{ignoregauge}
\ee
      where $a,f$ are spinor and flavour indices. $C^f_a $
      averages over a
      hypercube of sidelength $ a_0$ centered at $x$, and is only
      nonzero when $(a,f)$ matches with the pseudoflavour of $x$.
      Details will be given later on.
\item Admit dynamical gauge fields.\\
      The blockspin for the Higgs field will take the form
\be
   \Phi_{\alpha } (x) = \int_z C(U|x,z)\ \vp_{\alpha }(z)
\label{WHAT_WITH_GAUGE_AND_FERMIONS}
\ee
      similarly as in 1., but with an averaging kernel $C$ which
      depends on the gauge field $U$ for reasons of covariance.
      A  formula like~\eq{ignoregauge}
      with $U$-dependent kernels $C^f_a$
      will define the blocked Fermi fields. In
      addition one will need to define blocked gauge fields. This will
      be discussed later on.
\item Composite Higgs.\\
      There may be no fundamental Higgs field in the fundamental action.
      In this case one may try to define a composite Higgs field
      $\Phi $ in the effective action by a formula of the following kind
$$ \Phi_{\alpha} (x)= \int_z \int_w\sum
     \overline{\psi }^f_a\ C^{fg}_{ab}(U|x,z w)\ \lambda_{\alpha }
     \psi^g_b (w) \ . $$
     with matrices $\lambda_{\alpha }$ in colour space (colour indices
     of the Fermi fields $\psi $ are omitted).
\item Fully effective theory.\\
      In this case, the fields in the effective action are all
      interpolating fields for physical particles. They can be
      defined as composite fields similar to 4., but they will have
      to be gauge invariant. For instance, the physical electron field
      could take the form
   $$ \Psi (x)= \int_z\int_w \psi^f_a(z)\ C^f_a (U|x,zw)\ \vp (w) $$
     where $\psi^f_a$ are the left and right handed fundamental electron
     field and $\vp $ is the fundamental Higgs field.
\end{enumerate}
    One may wish to compute the ultimate effective action in a
    sequence of steps, lowering the cutoff in stages. In this case, one
  may have to proceed to
               different kinds of definitions of blockspins at
    different values of the UV-cutoff corresponding to different
    length scales of compositeness.
\section{The Constraint Effective Potential
           \label{SECTION_CEP}}

Starting from a fundamental theory on a finite volume, the
constraint effective potential is defined as the special case
of the effective action in which the lattice $\Lambda_0$ consists of
a single point. In this case, the block spins $\Phi $ may be
vectors in colour or flavour space, but they have no space time
argument x.

Consider for instance a fundamental theory which is a Higgs model
with or without dynamical gauge fields. For simplicity, we do not
include fermions. One introduces
$$ \Phi = \mbox{(gauge covariant) magnetization} $$
by a formula similar as before. The constraint effective potential
$V_{eff}(\Phi ) $ was introduced in
                  \cite{CEP_RWY} and is defined by
\be
   e^{-V_{eff}(\Phi )} = \int \D \vp\ \D U
   \ \delta (\Phi -C\vp )\ e^{-\H(U,\vp)}
  \label{DefCEP}
\ee
and similarly in the absence of gauge fields. C depends on U in the
presence of gauge fields.

     $\exp \left\{-V_{eff}(\Phi )\right\}$
gives us the probability density of the
magnetization $\Phi $. But
  $$
 V_{eff } \sim  \mbox{ volume } \es .
  $$
Therefore the minima of $V_{eff }$
become extremely sharp in the limit
of large volume, and in the infinite volume limit, the only possible
values which $\Phi $ may assume are absolute minima of $V_{eff} $.
Therefore, the shape of $V_{eff}$
in the large volume limit  gives information about
the phase structure of the theory. One has to look for minima.
If there are several absolute minima
\footnote{$\V_{eff}/\mbox{volume}$ becomes convex in the infinite
volume limit because the height of maxima of $V_{eff}$ between minima
increases less fast than the volume.}
       which are transformed into each other by a symmetry, then
that symmetry is spontaneously broken. For plots which show
the crossover of the effective potential from a single shape
(symmetric phase) to double well behaviour ( broken phase) see
Fig.~\ref{trans}.
\begin{figure}
 \vspace{20.0 cm}
\caption{Multigrid Monte Carlo data
 for $\lambda \varphi^{4}$-theory
 \protect\cite{EVALUATE_PHI4} with error bars
as a function of the magnetization $\Phi $ for bare coupling
$\lambda_u=16.376$
and fit with the one loop formula for the constraint
effective potential for three values of the bare mass
a) $m_u^2=-1.14$ , b) $m_u^2 =-1.15$ , c) $m_u^2 = -1.16$.}
\label{trans}
\end{figure}

If the height of different relative minima changes when one varies a %
parameter in the theory in such a way that the absolute minimum %
jumps, then the theory has a first order phase transition.
\section{Renormalization Group Picture
         \label{SECTION_RG_PICTURE}}
 In the renormalization group approach \cite{Wilson}
                                       one composes the blocking
 \ba
         \Lambda &\mapsto & \Lambda_0 \nn \\
         a &\mapsto & a_0    \nn \\
         \H &\mapsto & \H_{eff} \nn
 \ea
 from a sequence of blockings, lowering the cutoff in steps
 \ba
 \Lambda &\equiv & \Lambda_N \mapsto \Lambda_{N-1} \mapsto \dots
 \mapsto \Lambda_1 \mapsto \Lambda_0 \nn \\
 a &\equiv & a_N \mapsto a_{N-1} \mapsto \dots
 \mapsto a_1 \mapsto a_0 \nn \\
 \H &\equiv & \H^N \mapsto \H^{N-1} \mapsto \dots
 \mapsto \H^1 \mapsto \H^0 \equiv \H_{eff} \nn
 \ea
 where $a_{j-1}= La_j $ with $L=2,3 $ or so.

 The theory with action $\H^j $ lives on a lattice of lattice
 spacing $a_j$.  Considering one such theory at a time, one
 goes to the unit lattice by setting $a_j=1$. In this way one
 obtains a sequence of actions $ \tilde{\H}^j$.
 \be
 \tilde{\H}^N \mapsto \tilde{\H}^{N-1} \mapsto \dots
 \mapsto \tilde{\H}^1 \mapsto \tilde{\H}^0 \equiv \tilde{\H}_{eff} \nn
 \ee
These actions can be compared, since they all live on the same lattice
with lattice spacing $1$. This yields the
$$
\mbox{renormalization group flow: } \tilde{\H}^j\mapsto
         \tilde{\H}^{j-1}\es .
$$
Imagine that the actions are somehow parametrized. One distinguishes
between relevant, marginal, and irrelevant parameters. These parameters
are also called ``running coupling constants''.
\begin{description}
\item[ Relevant:]
A small change in a relevant coupling constant in $\H^j$
 changes $\H^{j-n}$ very much for large $n$.
\item[ Irrelevant:]
A small change in an irrelevant coupling constant
 changes $\H^{j-n}$ arbitrarily little for large $n$.
\item[ Marginal:]
in between.
\end{description}
In the limit of small $a$, many renormalization group steps will be
needed. As a result,
knowledge of all relevant and marginal coupling constants in $\H $
suffices to determine $\H_{eff}$. Theories with actions $\H $
whose relevant and marginal parameters all agree are said to belong to
the same universality class.

Let us remark that one can also lower an UV-cutoff (Pauli-Villars,
for instance) infinitesimally in one step. In this case the
renormalization group flow is determined by a functional differential
equation known as the renormalization group differential equation
\cite{RG_PICTURE_POLCHINSKI}.

In the conventional approach, all the actions $\H^j$ depend on the same
type of fields. But we noted  before that more flexibility
should be left. It may be necessary to change the kind of fields at
some length scales (of compositeness) $a_j$.
                          In particular, $\H^j $ may depend on
different kinds of fields than the fundamental action $\H $ when
$j < N$.

\subsection*{Effective observables}
In principle, expectation values
                                 $\langle O\rangle_{\H } $
     of arbitrary observables in the fundamental theory may be
translated into expectation values
$\langle O_{eff}\rangle_{\H_{eff}}$
  in the effective theory, such that
$$
   \langle O\rangle_{\H } = \langle O_{eff}\rangle_{\H_{eff}} \es .
$$
But this requires that one computes effective observables in
addition to the effective Hamiltonian \cite{EVALUATE_CARGESE},
\be  O_{eff}(\Phi ) = e^{\H_{eff}(\Phi )}
                           \int \D \vp\  O(\vp )\
     \delta (\Phi - C(\vp ) )\ e^{-\H (\vp )}\es %\label{DefHeff}
\es .
\ee
\section{Computation of Effective Actions by Perturbation Theory
         (Loop Expansion)
         \label{SECTION_EFFECTIVE}}
 for linear averaging maps $C$ in theories with or without gauge fields.
\subsection*{Scalar free field theory}
$$\H_{\rm free}(\vp )= \frac 12 \int_z \vp (z)[-\Delta +m^2]\vp (z) =
   \frac 12 (\vp , [-\Delta + m^2] \vp )\ . $$
 Herein, $\Delta $ may be either the ordinary  Laplacian, or the
 covariant Laplacian in some given external gauge field, and the
 original theory may live either on the continuum or on a lattice
 $\Lambda $. Its sites are denoted by $z,w$.
 Our aim is to compute an effective action on a lattice $\Lambda_0$
 of lattice spacing $a_0\leq O(m^{-1})$. Sites in $\Lambda_0 $ are
 denoted by $x,y$.

 We assume a linear averaging map $C$,
 $$\Phi (x)=\int_z C(x,z)\vp (z) \equiv C\vp (x)     \ . $$
 In the absence of a gauge field, we may choose
\be
   a_0^d\ C(z,x) =  \left \{ \begin{array}{ll}
    1 & \quad\mbox{if $z\in x$}\\
    0 & \quad\mbox{otherwise}
    \enspace  .         \end{array}
               \right.
\label{step}
\ee
 The original lattice $\Lambda $ of sites $z $ is covered by a block
 lattice $\Lambda_0 $, whose sites $x $ are blocks with side length
 $a_0 $, so we write $z \in x $ if the point $z $ is in the block $x$.
 More generally, $C$ may depend on an (external) gauge field.

 Following Kupiainen and Gawedzki
     \cite{WHAT_GK,EFF_ACT_PT_GAW_KUP},
 one splits the field $\vp $ into a low frequency part $\psi $,
 called the
           ``background field'', and a high frequency part $\zeta $,
 called the ``fluctuation field''. The low frequency field is
 supposed to be obtainable from the block spin $\Phi $ on $\Lambda_0$
 by interpolation with an interpolation kernel $\A $.
 \ba \vp (z) &=& \psi (z) + \zeta (z) \nn \\
             &=& \A \Phi (z) + \zeta (z) \ , \label{split}
  \ \ \mbox{with} \ \  \A \Phi (z)
                      = \int_{x\in \Lambda_0} \A (z,x)\Phi (x) \ .
 \ea
 If we demand that
 \be C\A = \bbbone  \ ,\label{CA} \ee
 then it follows that the fluctuation field has zero block average,
 \be C\zeta = 0 \nn \ . \ee

Now we insert the field split into $\H_{\rm free} $. Let us consider the
 case $m=0$. Generalization is obvious.
 \be
 (\vp ,\Delta \vp )= (\A \Phi , \Delta \A \Phi ) +
                     (\zeta , \Delta \zeta )+2(\zeta, \Delta \A \Phi )
  \ . \nn \ee
 We demand that $\A $ is chosen in such a way that the mixed term
 vanishes,
 \be (\zeta , \Delta \A \Phi )=0 \ . \label{mixed} \ee
 This gives
 \be
   \H_{\rm free} = \frac 12 (\Phi , -\Delta_{eff} \Phi ) +
                 \frac 12 (\zeta , \Delta \zeta ) \ ,
 \ee
 with
 \be \Delta_{eff } = \A^{\ast } \Delta \A \label{Deltaeff} \ . \ee
 Since $C\zeta =0$ it follows that the mixed term~\eq{mixed}
 vanishes as desired, provided
 \be -\Delta \A = C^{\ast }u^{-1} \label{Deltaeff'}\ee
 for some $u^{-1}$.

 Applying $\A^{\ast }$ to both sides of~\eq{Deltaeff'} it follows
 that
 \be u^{-1} = -\Delta_{eff}\ . \ee
 The operator  $\Delta_{eff}$ acts according to
 \be
 \Delta_{eff}f(x) = \int_{y\in \Lambda_0} \Delta_{eff }(x,y)f(y)
 \nn
 \ee
 with kernel
 \be
  \Delta_{eff } (y,x)= \int_z \A(z,y)^{\dagger }\Delta \A(z,x).
 \ee
 Here, $\Delta$ acts on the argument $z$,
 and $^\dagger $ denotes the hermitean conjugate of a matrix, or
 the transpose, if the matrix is real.

   \eq{Deltaeff'} becomes
 \be \Delta \A (z,x) = \int_y C(y,z)^{\dagger } \Delta_{eff}(y,x) \ .
 \label{Deltaeff''} \ee
 If $C(x,z)$ is the step function~\eq{step} then~\eq{Deltaeff''}
 says that $\Delta \A (z,x) $ should be constant on blocks $y$ as a
 function of $z$. The kernels $\A $ and $\Delta_{eff}$ are determined as
 solutions of \eq{Deltaeff''} together with the subsidiary condition
 \eq{CA}.

 {\bf Remark:} The field
 $\psi (z) = \A \Phi (z) $ minimizes
  $\H_{\rm free }(\psi )$
 subject to the subsidiary condition that
 $\psi $ has the prescribed block average $\Phi $, viz.
 $$
   C\psi (x)= \Phi (x) \es .
 $$
  This remark can be used to compute the kernel $\A$ by stan\-dard
 op\-ti\-mi\-za\-tion al\-go\-rithms.

 Numerical results confirm that the kernel $\A (z,x)$ decays
 exponentially with the distance of $z$ from the center of the block
 $x$, with decay length $a_0$. This is true with and without an
 external gauge field in $\Delta $, see for instance
 Fig.~\ref{FigBoseA} of Sect.~\ref{SECTION_PROPAGATOR}.
 As a result, $\Delta_{eff}(x,y)$ decays exponentially as a function of
 $|x-y|$ with decay length one lattice spacing $a_0$.

 Now we are ready to compute the effective action. By definition
 \be
     e^{-\H_{eff }(\Phi )} = \int \D \vp \delta (\Phi -C\vp )
     e^{-\H_{\rm free }(\vp )}  \es .
 \nn
 \ee
 We shift the field $\vp $ by an amount depending on $\Phi $,
 using $\zeta = \vp - \A \Phi $ as a new variable of integration.
 Since $C\A =\bbbone $ by assumption, it follows that
 \be
 e^{-\H_{eff }(\Phi )} =
 e^{\eh(\Phi , \Delta_{eff} \Phi )}
     \int \D\zeta\ \delta (C\zeta )\ e^{-\H_{\rm free }(\zeta )} \es .
 \ee
 The $\zeta $-integral merely produces a constant independent of $\Phi$.
 Therefore
 \be \H_{eff}(\Phi ) = \eh (\Phi , -\Delta_{eff }\Phi ) +
     \mbox{ const. }
 \es  .
 \ee
 $\H_{eff}$ has good locality properties because of the aforementioned
 decay properties of $\Delta_{eff}(x,y)$. There are exponential tails,
 but this is tolerable and in general inevitable.
 \subsection*{Split of the propagator}
 The split of the field $\vp $ induces a split of the free propagator
 $ v = (-\Delta )^{-1}$ into a low frequency propagator
 $ v_{low} $ and a high frequency propagator  (fluctuation field
 propagator) $\Gamma $. For proper choice of the
 normalization factor $\N $,
 \be
  d\mu_{\Gamma }(\zeta ) =
  \N^{-1} \D \zeta\ e^{-(\zeta , -\Delta \zeta )/2 }\ \delta (C\zeta )
 \ee
 is a normalized Gaussian measure with covariance
 $$
  \G (z,w) = \N^{-1} \int \D \zeta\  \delta (C \zeta )\
  e^{-(\zeta , -\Delta \zeta )/2}\ \zeta (z)\ \zeta (w)
 \es .
 $$
 Thinking of the $\delta$-function as a limit of a Gaussian, one sees
 that
 \be
      \Gamma = \lim_{\kappa \mapsto \infty }
        (-\Delta + \kappa C^{\ast }C )^{-1}
 \label{Gamma}
 \ee
 Since $\zeta $ is supposed to be the high frequency part of the
 field $\vp $, it is a field with an infrared cutoff (equal to the
 UV-cutoff of the effective theory). Therefore one expects that
 $\Gamma (z,w)$ decays exponentially with the distance of $z$ from $w$,
 with decay length $a_0$. Analytical and numerical results confirm that
 this is indeed the case. This remains true when there is a
                           a gauge field in
 $\Delta $, provided the kernel $C$ is chosen in the right way.
 A proper choice will be introduced later.

 Let us now compute the 2-point function.
 \ba
   \langle \vp (z)\vp (w)\rangle
   &\propto &\int \D \Phi\ \D \vp\ \delta (\Phi -C\vp )\
   e^{-\H_{\rm free}(\vp)}\ \vp (z)\ \vp (w) \\
   &=& \int \D \Phi\ \D \zeta\  e^{(\Phi ,-\Delta_{eff}\Phi )
       + (\zeta ,-\Delta \zeta )}
   \ \delta (C\zeta )
                 \bigl( \A \Phi (z)+\zeta (z)\bigr)
                 \bigl( \A \Phi (w)+\zeta (w)\bigr)  \ea
Since $u = (-\Delta_{eff })^{-1} $ it follows that
$$ \N_{\Phi} \cdot\ \int
 \ \D \Phi \ e^{(\Phi ,-\Delta_{eff}\Phi )/2}
 \ \Phi (x)\ \Phi (y) = u(x,y)       $$
for a proper choice of the normalization factor $\N_{\Phi}$.\newline
As a result we get the following final result for
$\langle\vp (z)\vp (w)\rangle\equiv v(z,w) $
\be v(z,w) = v_{low}(z,w) + \Gamma (z,w) \ ,\label{splv} \ee
with
\be v_{low } = \A \ u\  \A^{\ast }\ \ , \ \ u=-\Delta_{eff}^{-1} \ . \ee
More explicitly, \eq{splv} reads
  \ba
   \G(z,w)&=& v(z,w)-\int\int_{x,y\in\L^1}
   \A(z,x)u(x,y)\A^{\dagger }(w,y) \es  \label{gamm}\\
    v &=& -\Delta^{-1} \ \ \mbox{and } \ \ u=-\Delta_{eff}^{-1} \es
   \label{GammaExp}\ea

\subsection*{Perturbation theory}

We will now show how to compute effective actions by perturbation
theory.
To be specific, let us consider an action for a scalar field $\vp $ of
the form
\be
  \H (\vp )= \int_z \bigl\{ -\eh \vp (z)\Delta \vp (z) +
        \V (\vp (z) )\bigr\} = (\H_{\rm free} + V)(\vp )\ . \ee
 with
 \be \V (\xi )= \eh m_{u}\  \xi^2 + \frac {1}{4!}\  \lambda_{u}\
 \xi^{4} \ , \ee
 for instance.
 Introduce
 \be V_{\Phi }(\zeta ) = \int_z \V \bigr(\A \Phi (z)+\zeta (z)\bigl)
 \ . \ee
 This determines a self interaction of the high frequency field $\zeta$
 which depends parametrically on the block spin $\Phi $.

 Proceeding as before one finds the following expression for the
 effective action
 \be \H_{eff } (\Phi ) = -\frac 12 (\Phi , \-\Delta_{eff }\Phi )
  + \hat{V}_{eff}(\Phi ) \label{HeffPT}
  \ee
 with
 \be
 e^{-\hat{V}_{eff}(\Phi )}= \int d\mu_{\Gamma }\ e^{-V_{\Phi }(\zeta )}
 \es . \label{VF} \ee
 We remember that $d\mu_{\Gamma } $ is the normalized
                                           free field measure with
 propagator $\Gamma $.  The logarithm of the $\zeta $-integral
 can be computed by standard perturbation theory. It is given
 by a sum of Feynman diagrams, whose lines represent propagators
 $\Gamma $, and whose vertices depend parametrically on $\Phi $.

 The resulting effective action $\H_{eff}$ will have good locality
 properties, provided the fluctuation field propagator
 $\Gamma (z,w)$ and the interpolation kernel $\A (x,z)$ decay
 exponentially with decay length $a_0$.

 Perturbation expansions of the kind described here were first done by
 Kupiainen and Gawedzki
     \cite{WHAT_GK,EFF_ACT_PT_GAW_KUP},
 as part of a rigorous renormalization group analysis.
 Constraint effective potentials were also computed by another method
 in \cite{Wetterich}.
 \subsection*{Constraint effective potential}
 We specialize the results of the previous subsection to the
                                           case of the constraint
 effective potential. In this case,
 $\Lambda_0$ has only one point $x$. Therefore the kernel of
 $\Delta_{eff } $ can only have $(x,x)$ as its only pair of variables.
 We admit the possibility that $\Delta $ is the covariant Laplacian
 in an external gauge field. Of course, $\Delta_{eff }$ will also
 depend on $U$ in this case. We abbreviate
  $$ \Delta_{eff}(x,x)= \epsilon_{0} \ . $$
 Eq.~\eq{HeffPT} simplifies to
 \be
  \H_{eff}(\Phi )\equiv V_{eff}(\Phi ) =
        \Phi\ \epsilon_{0}\  \Phi + \hat{V}_{eff}(\Phi )
  \es .
  \label{CEFF}
  \ee
 with $\hat{V}_{eff}$ from~\eq{VF}. \newline
 In the presence of an external
    gauge field $U$, $\epsilon_{0}$ depends on $U$.
                    $\hat{V}_{eff }$ will also depend on $U$,
 because the high frequency propagator $\Gamma $ does, and the kernel
 $\A $ as well.                                        In addition
 there is a gauge field dependence in the normalization factors of the
 Gaussian measures. This leads to a $\Phi $-independent, but
 $U$-dependent term in $\hat{V}_{eff} $ which will be ignored for
 now, but which has to be taken into account later on when we consider
 dynamical gauge fields.

 Without gauge field, $\epsilon_{0} =0$.
 But in the presence of a gauge
 field which is not a pure gauge, $-\Delta $ is strictly positive,
 and therefore $\epsilon_{0} $ is also strictly positive.

 Let us emphasize that $\H_{eff }$ has no infrared problems because
 $\Gamma $ has a built-in infrared cutoff.

 \subsection*{1-loop approximation}

 In 1-loop approximation, the contribution $\hat{V}_{eff }$ to the
 constraint effective potential~\eq{CEFF} becomes % \cite{ PALMA}
\cite{EFF_ACT_PT_PALMA_1,FP2,BP2}
 \be
 \hat{V}_{eff}(\Phi ) =
   \int \int_{z,w} \V '(\A \Phi (z))\Gamma (z,w)
                   \V '(\A \Phi (w) )
 + \eh \Tr\ \ln (-\Delta +\V'' (\A \Phi ) )  \label{1loop}\ee
 where a prime
           $(^\p)$ indicates a derivative with respect to the argument,
 and
 $\V '' $ is the multiplication operator defined by
 \be (\V '' (\A \Phi )f )(z) = \V '' (\A \Phi (z))f(z) \nn \ . \ee
 In graphical notation
 \bdm
  \Tr\ \ln (-\Delta +\V'' (\A \Phi ) ) &=& \sum_{n\geq 1}
\frac {(-)^n}{n} \int_{z_1}\dots \int_{z_n}
 \edm
with Feynman rules
$$ z \ \_\_\_\_\_\_\_\_\_\_\_\_\_\_\_ \ w = \Gamma(z,w) \ \ \
, \ \ \ \ \ \ \ \ \ \ \ \ \ \ \ \otimes_z =
                     \V '' (\A \Phi (z)) \ . $$
The first  term in expression~\eq{1loop} is not 1-particle
irreducible. This term vanishes in the absence of gauge fields, and more
generally it vanishes up to second order in the gauge coupling constant.

The contribution $\hat{V}_{eff } $ to the full effective
action~\eq{HeffPT} on a lattice $\Lambda_0$
                                     of more than one point has the
same form. The 1-particle reducible contribution is important in this
case \cite{EFF_ACT_PT_GAW_KUP}.

\subsection*{Use of the 1-loop formula to compute finite size
corrections to the constraint effective potential}

For definiteness, let us consider $\lambda\vp^4$-theory.

 {}From the constraint effective potential/volume in the infinite volume
limit one can extract renormalized coupling constants, provided one
knows the wave function renormalization constant from an independent
calculation.

In fact, according to the standard renormalization of the
 $\lambda\vp^4$-theory, one needs to define the three quantities
 $ m_{r}, \lambda_{r} $ and $ Z_{\vp} $ ( renormalized mass,
 self-coupling constant and wave function renormalization constant
respectively).
This can be achieved in the symmetric phase through the second and
the fourth derivatives of the constraint effective potential at
its minimum, viz.

\be
    m^{2}_{r} = \frac{U''_{eff} (\Phi_{\rm min})}{Z^{-1}_{\vp}}
\label{veff2ndderiv}
\ee

and for the self-coupling constant

\be
   \lambda_{r} = \frac{U''''_{eff} (\Phi_{\rm min})}{Z^{-2}_{\vp}}
\label{veff4ndderiv} .
\ee

In the broken phase it is useful to define the renormalized square
mass also through~\eq{veff2ndderiv} but the self-coupling
constant $\,\lambda_{r}\,$ is now defined
  through the tree level relation:
 \be  \lambda_{r} = 3\, m^{2}_{r}\, / v^{2}_{r} \label{3level}  \ee
  where $\,v_{r}\,$ is the vacuum expectation value of the
  field $\,\vp_{r} = Z^{- 1/2}_{\vp} \vp$ and $U_{eff}= V_{eff}$/volume.

 One can obtain a renormalized loop expansion for the constraint
 effective potential by  manipulating the bare expansion.
 Eqs.~\eq{veff2ndderiv}~-~\eq{3level} determine the
 renormalized coupling constants as functions of the bare
 parameters. Inverting this relations,
 one obtains the bare quantities as a function of the
 renormalized ones. One may use them to express $\V$ in
 expression~\eq{1loop} ( or its corresponding n-loop generalization )
for the constraint effective potential as a function of the renormalized
 coupling constants. Reordering the resulting expansion in number of
 loops gives the desired expansion in terms of
 $\lambda_{r}, m_{r}$.

In practice, numerical computation of constraint effective potentials is
only feasible
                     on finite and not too large lattices.

The following strategy, which was first proposed by one of us (G.P.)
in \cite{EFF_ACT_PT_PALMA_1}, eliminates the need to use very large
  lattices.
It makes it possible to extract renormalized coupling constants
 from numerical data for values of the bare coupling constants which are
very close to the critical line,
                               and for lattice sizes which need not be
large compared to the correlation length.

Triviality of $\lambda \vp^4$-theory implies that the renormalized
coupling constant $\lambda_r $ is necessarily small when one is close
to a critical point, provided the UV-cutoff is high enough, i.e.
provided $a \ll$ correlation length.

Therefore, a perturbative calculation of the constraint effective
potential on a {\it finite lattice} { \rm as a function of the
renormalized coupling constants} will yield accurate results.
In the practical application \cite{EFF_ACT_PT_PALMA_1,FP2}
it was verified that the
two loop corrections are indeed very small.

The result of this computation is fitted to numerical data taken at
various values of the bare coupling constants to extract the
renormalized coupling constants as functions of the bare ones.

Some results of this kind are shown below.
Numerical results of  Meyer and Mack \cite{EVALUATE_PHI4}
are shown in Fig.~\ref{trans}
together with a fit by a renormalized 1-loop formula for the
constraint effective potential. Numerical results for the constraint
effective potential for 4-component $\varphi^4$-theory at stronger
coupling but further away from the critical line were presented in
\cite{EVALUATE_KUTI}.

The results are compared with analytical results of
L\" uscher and Weisz \cite{EFF_ACT_PT_LU_WEI}
in Table~\ref{comp} .

\begin{table}
\caption{Renormalized parameters from the fits of
 Fig.~\protect\ref{trans}
 and the corresponding values obtained from
 L\"uscher and Weisz ($^{LW}$).}
\label{comp}
 \begin{center}
    \begin{tabular}{|l||c|c|c|c|} \hline
    &$m_{r}\,(\infty)$&$\lambda_{r}%
    \,(\infty)$&$m_{r}^{LW}$&$\lambda_{r}^{LW}$\\ \hline\hline
    $m^{2}_{0} = - 1.14$&$%
                 0.079\pm 0.003$&$7.5\pm 0.6$&$0.08$&$7.4\pm 0.2$%
                 \\\hline
    $m^{2}_{0} = - 1.15$&$0.06\pm 0.005$%
                 &$7.1\pm 0.1$&$0.06$&$7.2\pm 0.2$\\\hline
    $m^{2}_{0} = - 1.16$&$0.13\pm 0.008$&$%
                 7.8\pm 0.1$&$0.13$&$7.6\pm 0.2$\\ \hline
    \end{tabular}
\end{center}
\end{table}

Perfect agreement is found.
In this way, the nonperturbative part of the analytical computation
of L\"uscher and Weisz is sucessfully tested.
\section{Averaging and Interpolation Kernels $C$ and $\A $ in the
Presence of Gauge Fields, for Bosons \label{SECTION_BOSONS}}
In the presence of a gauge field $U$, the definition of a blockspin
$\Phi $ for a
  ($n$-component) scalar field $\vp  $ and the split of this field into
low frequency and high frequency part read
\ba
\Phi &=& C(U)\vp \ , \nn \\
\vp &=& \A (U) \Phi + \zeta \ , \nn \\
 C(U)\A (U) &=& \bbbone \es . \label{subs} \ea
 The operators  $C$ and $\A $ have kernels
 \ba
\A(U)\Phi (z)&=& \int_{x\in \Lambda_0}\A (U|z,x)\ \Phi (x) \ , \\
         C(U)\vp (z)&=& \int_{z\in \Lambda} C (U|x,z)\ \vp  (z)
  \es .
  \ea
As usual we denote points of the block lattice $\L^0$ by $x,y$ and
points of the fundamental lattice $\Lambda $ by $z,w$.

In the absence of gauge fields, a good choice of $C$ is given by
\ba
   C^\ast (z,x) \equiv C(x,z)^\dagger =
               \left \{ \begin{array}{ll}
    1 a_0^{-d} &\, \mbox{ if $z\in x$ }\nn\\
    0 \quad    &\, \mbox{ otherwise }
    \es ,               \end{array}
               \right.
\ea
($^{\dagger }$ denotes the hermitean conjugate of a matrix.)
 This satisfies
 \be
   -\Delta_{N,x}C^{\ast } (z,x) =
               \left \{ \begin{array}{ll}
   \epsilon_0(x)\ C^{\ast }(z,x)
      &\, \mbox{ if $z\in x$ }\nn\\
    0 &\, \mbox{ otherwise }
    \es .               \end{array}
               \right.
\label{EVC}
\ee
 $\Lambda_{N,x}$ is the lattice version of the Laplacian with Neumann
 boundary conditions on the boundary of the block $x$,
 and $\epsilon_0(x)= 0 $ is its lowest eigenvalue. In the presence of a
 gauge field, the Laplacian with Neumann boundary conditions takes the
 form
 \be \Delta_{N,x}f(z)= a^{-2}
 \sum_{w\ nn. z, w\in x} [ U(z,w)f(w)-f(z)]   \es .
 \label{LapNbc}
 \ee
 $U(z,w)$ is the lattice gauge field attached to the link from $w$ to
 $z$, the summation runs over those nearest neighbours $w$ of $z$
 which are inside the block $x$.

 {\bf Proposal:}
 \cite{KalNP}
               Retain the eigenvalue equation~\eq{EVC} for
 $C$                                                        in the
 presence of gauge fields, with covariant Laplacian $\Delta_{N,x}$
 as defined in~\eq{LapNbc}, and $\epsilon_0(x)$ equal to its lowest
 eigenvalue.
 Of course, both $C$ and $\epsilon_0(x)$ will depend on the gauge field.

 The interpolation kernel $\A $ may also be defined by the same equation
 as before. There should be $u(y,x)$ such that
 \be -\Lap \A (z,x) = \int C^{\ast }(z,y)\ u(y,x) \es . \ee
 Here $\Delta $ is the gauge covariant Laplacian on the whole lattice.
  $u$ is a Lagrange multiplier
 which is determined by the subsidiary condition~\eq{subs}.

 In the special case when the block lattice
                          $\Lambda_0 $ consists of a single point $x$,
 the Neumann boundary conditions are inoperative, because the block
 $x$ has no boundary when periodic boundary conditions for $\Lambda $
 are in force. Therefore $\Delta_{N,x}=\Delta $. The equation for $\A $
 is therefore  satisfied by
 \be \A = C^{\ast }  \quad
     \mbox{if } \Lambda_0 = \mbox{ single site } \es .
 \ee
 Let us
 next discuss the arbitrariness in the solution of~\eq{EVC}.
 Consider for example the gauge group $G=SU(2)$. For fixed $x$,
 the $2\times 2$ matrix $C(z,x)$ may be thought to be composed of
 two 2-vectors, each of which is a solution of the
 eigenvalue equation~\eq{EVC}. (It can be proven that there exist two
 linearly independent solutions with the same eigenvalue.)
 Arbitrary linear combinations of these two solutions will also be
 solutions. This reflects itself in the arbitrariness
 \be C^{\ast }(z,x) \mapsto C^{\ast }(z,x)\ \Omega (x)\ee
 of the $2\times 2$-matrix solution $C$ of~\eq{EVC}, where $\Omega (x)$
 are arbitrary $2\times 2$-matrices. To eliminate this arbitrariness,
 we may impose subsidiary conditions
 \begin{enumerate}
 \item Normalization: $CC^{\ast }=\bbbone $.\\
    This restricts the arbitrariness to unitary matrices $\Omega (x)$.
 \item Gauge transformation property:\\
  Select a distinguished site $\hat{x}$ in the block $x$ -- for instance
  the center of the block -- and demand that
  $C^{\ast }(\hat{x},x) > 0 $ as a matrix. This makes the kernel $C$
  unique. It will transform under a gauge transformation
  \footnote{
    Re\-mem\-ber that un\-der a gauge trans\-for\-mation
  $ \vp(z)\mapsto g(z)\vp (z) $ and
$U(z,w)\mapsto g(z)U(z,w)g(w)^{-1}$.
The covariant Laplacian $\Delta_{N,x}$ is gauge covariant.
  Therefore the eigenvalue equation~\eq{EVC}
  is also gauge covariant.} %
  according to
  \be C^{\ast }(z,x)\mapsto g(z)C^{\ast }(z,x)g(\hat{x})^{-1}  \ .
  \label{Cgauge} \ee
 \end{enumerate}
  It follows that the blockspin $\Phi (x) $ transforms like a field
  sitting at the block center $\hat{x}$,
  \be \Phi (x)\mapsto g(\hat{x})\Phi (x) \ . \ee
  Given $C$, the kernel $\A $ is also unique and it transforms under
  gauge transformations exactly like $C^{\ast} $.

  There exists an efficient algorithm for computing the gauge field
  dependent kernels $C$ and $\A $ \cite{KalNP}.

  Let us consider the implications for the effective potential
  in the presence of dynamical gauge fields,~\eq{DefCEP}.
  Since the U-dependent kernel $C$ will depend on the choice of
  $\hat{x}$, the gauge covariant magnetization will also depend on
  $\hat{x}$. But it is readily verified that the effective
  potential~\eq{DefCEP}
  is independent of the choice of $\hat{x}$
  \cite{BP2}.
\section{Effective Potential for a Higgs Model with Dynamical Gauge
 Fields\label{EFFECTIVE}}
 Let us consider a Higgs model with dynamical gauge fields but
                               without fermions, with a gauge group
 $SU(2)$ and a 2-component scalar field $\vp $. Let

 \be  \Phi = \mbox{gauge covariant magnetization} = \int_zC(z)\vp(z)
  \ , \nn \label{cov_magne}  \ee
 defined with the help of the gauge covariant kernel $C(x,z)$ which
 was described in the last section. The argument $x$ is redundant
 because there will be only one point $x$ in $\Lambda_0$.
 Of course, $C$ and $\Phi $ will depend on the gauge field $U$, and
 so does $\A = C^{\ast }$.  By definition,
 \ba e^{-V_{eff }(\Phi )} &=& \int \D U \ e^{-\tilde{V}(U,\Phi )} \nn \\
     e^{-\tilde{V }(U,\Phi )}&=& \int \D \vp\
     \delta (\Phi - C\vp )\ e^{-\H (U,\vp )} \ . \label{DefCEP'} \ea
 $\H $ is the standard action for a Higgs model. Using an
 appropriate lattice definition of the field strength tensor
 $F_{\mu \nu }$ \cite{MacDLGT},
the Wilson action for the lattice gauge field can
 be written in standard $ F_{\mu \nu }  F_{\mu \nu }$-form, and
 \ba \H &=& \int_z \left [ \frac 14
 F_{\mu \nu }  F_{\mu \nu }^{\ast }
 + \frac 12 \langle \vp , -\Delta  \vp \rangle
                                        + \V (\vp (z))\right ] \nn \\
   \V (\vp (z))&=&  \frac 12 m_u^2\vp^2(z) +
 \frac{1}{4!}\lambda_u \vp^4(z) \ . \label{SU2H_ACTION}\ea
 $\Delta $ is the gauge covariant Laplacian. It depends on $U$.

 The functional integral will be performed in two steps.
(For details of the loop expansion, see \cite{BP2}.) %
 Calculations by a different method were reported in %
 \cite{CEP_WETTERICH} for a $U(1)$-gauge field.) %
  \newline \newline%
 {\large{\bf{$\vp $ integration}}} \newline \newline%
 $\D\vp\ \delta (\Phi - \int_z C(z)\vp (z) )$ is an unnormalized
 Gaussian measure with nonzero mean. First we shift the field to
 bring the mean to zero. We introduce a new integration variable
 $\zeta $ in place of $\vp $,
 \be \vp (z) = \eta (z) + \zeta (z)  \ \ \ \mbox{with} \ \ \
 \eta (z)= \A (z)\ \Phi \label{shift'}\ee
 Since $\Lambda_{0}$ is a single point, the choice $\A = C^{\ast}$
 is in agreement with previous definitions.
 The interpolation kernel satisfies therefore the
 eigenvalue equation
 \be -\Lap \A (z) =  \epsilon_0 \, \A (z) \ ,
 \label{A}
 \ee
 $\epsilon_0 $ = lowest eigenvalue, $\Delta$ = covariant Laplacian
 on the lattice with periodic boundary conditions.

 Using $\A^{\ast }\A = C\A  = \bbbone $, the kinetic term becomes
 \be
  \langle
     \vp ,\  -\Delta \vp \rangle
                          = \Phi \epsilon_0 \Phi %\cdot \mbox{ volume }
  + \langle\zeta , -\Delta \zeta \rangle \nn
 \label{kinetic}
\ee
  so that
 \begin{equation}
  e^{-\tilde{V}(U,\Phi )} = e^{-\eh \Phi \epsilon_{0} (U) \Phi }
  \ \int \D \zeta \ \delta (\int_z C(z)\zeta (z) )
  \quad\exp \left( -\eh \langle \zeta, -\Delta \zeta \rangle
 + \int_{z} \V (\eta (z)+
  \zeta (z) \right) \es .  \end{equation}
 The eigenvalue problem~\eq{A} can be solved by time independent
 perturbation theory which is familiar from elementary quantum
 mechanics. Set
 \be U(z+ e_{\mu }, z) = 1 - g A_{\mu }(z) + \dots
 \es .
 \ee
 To order $g^2$ one finds
 \be \epsilon_0 = \frac{g^2}{8}  \int_z |A_{\mu }(z)|^2 + \frac{g^2}{4}
 \ds_{\mu,\nu} \int_z\int_w A_{\mu }(z)\Gamma (z,w) A_{\nu } (w) \ .
 \label{g2}
 \ee
 The high frequency propagator $\Gamma $ is only needed
 to zeroth order in $g$ and the expression valid  in a scalar
 theory without gauge fields can therefore be used for it at this point.

 In conclusion
 \be \int \left\{ \frac 14 F_{\mu \nu }F_{\mu \nu }^{\ast }
          +\frac 12 \Phi \epsilon_0 \Phi\right\}
 =   \int \frac 14 F_{\mu \nu }F_{\mu \nu }^{\ast }
   + \eh m^2(\Phi )\int |A_{\mu }|^2  + \dots.
\nn
\ee
   with
 \be m^2(\Phi ) = \frac 14 g^2 |\Phi |^2   \ .\label{MSQ} \ee
 We see that a $\Phi $-dependent mass term
 for the gauge field has appeared.

In the full theory, the fluctuations of the random variable $\Phi $
tend to zero in the infinite volume limit (cp. J. Klauders lectures at
this school \cite{HIGGS_KLAUDER}),
and its value becomes fixed to the position
$\hat{\Phi }$ of an absolute minimum of $V_{eff}(\Phi )$. As a result
\be m^2(\hat{\Phi })= \mbox{W-meson mass}   \ . \ee
The integral
\be \int \D \zeta \exp \left( -\frac 12 \langle
                                        \zeta, -\Delta \zeta \rangle
      +  \int \V (\eta (z)+ \zeta (z))\right) \label{int} \ee
is computed by expanding the exponent in the gauge coupling constant
$g$ and using the standard formula for a Gaussian integral

\begin{equation}
 \int \D \zeta\exp \left\{
 - {\ds_{z}} \left\{ \frac{1}{2} \langle \zeta | M_{z} \zeta \rangle
 + \langle \zeta | J_{z} \rangle \right\}
  \right\}  =
              (\det M)^{1/2} \exp \left( \frac%
 {1}{2}\sum_{z} \langle J_{z} | M^{-1}_{z} J_{z} \rangle \right)
\label{Gauss_Int}
\es .
\end{equation}

Expressing it in terms of the  {\it normalized} { \rm Gaussian
measure} $d\mu_{\Gamma }$ with covariance
(= free propagator) $\Gamma $, the integral~\eq{int} equals
\be  e^{-\frac 12 \Tr_C\ \ln \Gamma} \int d\mu_{\Gamma }(\zeta )
 \ e^{-\int_z \V (\eta (z) + \zeta (z) )} \ .
\ee
The high frequency propagator $\Gamma $ depends on the gauge field,
and so does $\A = C^{\ast} $ and therefore
            $\eta $,
                    %$\eta $ was
defined in~\eq{shift'}.
$\Gamma$ was defined in~\eq{Gamma} and
$$ \Tr_C = \mbox{trace on the space of functions } f \mbox{ which obey }
         Cf = 0 \ . $$
With some effort, $\Tr_C\ \ln \Gamma $ can be
                                             evaluated to second order %
in $g$. It is a quadratic expression in the gauge fields which %
will add to the kinetic term of the gauge field.%
\newline \newline%
{\large{\bf{Gauge field integration}}}\newline%
\newline%
Expanding everything to second order in the gauge coupling constant $g$,
the integration over the gauge fields becomes a Gaussian integration
which may be performed. Thanks to term $\Phi \epsilon_{0} \Phi$ in
in~\eq{kinetic}
and also additional quadratic terms in the vector
potential $A_{\mu }$ which resulted from the $\vp $ integration,
the kinetic term for the gauge field is nondegenerate.

In fact, it turns out that the whole quadratic contribution in the
gauge field $A_{\mu}$ in the large volume limit is equal to

\begin{equation}
 S^{q}_{g} [A] =  \frac{1}{2 v}\,{\ds_{k}}\,{\ds_{\mu \nu}}\,
 \delta_{r s}
              \left\{ \delta_{\mu \nu} [\hat{k}^{2} + m^{2} (\Phi)
 ] - \left( 1 + \frac{m^{2} (\Phi)}
 {\hat{k}^{2}}\, (\Phi)%
 \Theta\,(\hat{k}^{2}) \right) \, \hat{k}_{\mu}\,\hat{k}_{\nu} \right\}
 |\tilde{A}_{\mu }(k)|^{2}
\label{quadoper}
\end{equation}
 where $\tilde{A}_{\mu }(k)$ is the Fourier transform of the gauge
 field $A_{\mu}(z)$ and $\Theta (\hat{k}^{2})$ is
 equal to one for $\hat{k}^{2} \neq 0$ and the interpolation to conti%
 nuous $\hat{k}^{2}$ vanishes faster than $\hat{k}^{2}$ for $\hat{k}%
 ^{2} \rightarrow 0$.

The quadratic operator of~\eq{quadoper} {\it does} have an
inverse (the propagator) and therefore no gauge fixing is necessary!

The final result in finite volume is given in~\eq{veff-vol} in
the Appendix.

 Comparison with known results \cite{HIGGS_MONT}
 for the conventional effective
 potential reveals that \eq{veff-vol} simplifies
 in the infinite volume limit to
\be V_{eff} (\Phi )/ \mbox{ volume }
   = \mbox{conventional effective potential in the Landau gauge.}  \ee

The conventional effective potential is the Legendre transform of the
generating function for Greens functions, with a constant field as its
argument. It has been computed before~\cite{HIGGS_MONT}
on a finite
lattice $ L^{4}$ in the Landau gauge.

 If one performs a formal continuum limit in addition to the infinite
 volume limit in order to replace sums over k by integrals,
 the expression~\eq{veff-vol}
 simplifies (after inserting  counterterms into the action)
 to the following expression for the
 one-loop constraint effective potential $U_{eff} = V_{eff}/\,{\rm
 volume}$
  \cite{BP2}.

 \be
   \begin{array}{ll}
U_{ef\!f}(\Phi) =& U_{cl} (\Phi) + \frac{1}{64 \pi^{2}}\,\left\{%
   \left( U''_{cl} (\Phi)\right)^{2} \left[ \ln \left( U''_{cl} (\Phi)%
   \right) - \frac{1}{2} \right] \right.   \\
   \\
  &+ 3 \left( \frac{U'_{cl}}{\Phi}\,(\Phi)%
   \right)^{2}\,\left[ \ln \left(\frac{U'_{cl} (\Phi)}{\Phi}%
    \right) - \frac{1}{2} \right]   %
    + 9 \mu^{4}_{W} (\Phi)%
 \left.\left[ \ln \left( \mu^{2}_{W}%
    (\Phi) \right) - \frac{1}{2}\right]%
   \right\}
   \label{veff_vinf}
   \end{array}
  \ee

  with

   \be
         U_{cl} (\Phi) =    %
  \frac{1}{2}\,m^{2}_{r}\,\Phi^{2} + \frac{1}{4!}%
  \,\lambda_{r}\, \Phi^{4}\ \ \ {\rm and}\ %
  \ \ m^{2}(\Phi) = \frac{1}{4}%
  \,g^{2}_{r}\, \Phi^{2}.
  \ee

The expression~\eq{veff_vinf} is manifest gauge invariant and it
agrees with the known result
of Anna Hasenfratz and Peter Hasenfratz \cite{HIGGS_HASEN}
for the conventional effective potential computed
in the Landau gauge. $\Phi $ is the unrenormalized covariant
magnetization, i.e. no  wave function renormalization factor is
included in its definition.

Small renormalized couplings $\lambda_r = O(g^4_r)$ are within the
domain of validity of the perturbative result of~\eq{veff_vinf}.
Its shape is shown in Fig.~\ref{1st-trans}.
\begin{figure}
\vspace{8.5 cm}
\caption{The Constraint Effective Potential in the presence of
gauge fields in the infinite and large cutoff limit.
The gauge coupling
was fixed to be $g^{2}_{r} = 0.4$, the renormalized mass
$m^{2}_{r}$ is given by
$ m^{2}_{r} = - \frac{2}{3}\,\frac{\lambda_{r}}{g_{r}^{2}}$
and the scalar self coupling constant assumes values on
both sides of the critical line (solid line).
We observe that the minimum jumps from zero to $\Phi_{\rm min}$
when the parameters are varied from the symmetric to the Higgs phase.
This means that the system has a weak first order transition.}
\label{1st-trans}
\end{figure}
Inspection of Fig.~\ref{1st-trans}
shows that the absolute minimum jumps when the
parameters are varied. Therefore there is a first order phase
transition at weak couplings. This was noted before by
Hasenfratz and Hasenfratz \cite{HIGGS_HASEN},
following the pioneering work of Coleman
and Weinberg \cite{HIGGS_COLE_WEIN} .
\section{Blockspin for Fermions
         \label{SECTION_FERMIONS}}
As already discussed in
T. DeGrand's lecture \cite{FERMIONS_DEGRAND},
the naive
discretization of the continuum fermion action results in a
lattice model
with unphysical low energy modes, called ``doublers''. To reduce
the number of flavours, i.e. the degeneracy
of the lattice Dirac operator, one can use one-component
``staggered''
fields which live on different lattice sites.  In $d$ dimensions, a
staggered fermion field describes physical fermions with $2^{[d/2]}$
flavours.

We shall first be concerned with blockspins of the
form~\eq{ignoregauge}, i.e. we ignore
   gauge fields for now and consider only fermions.
\subsection*{Free Staggered Fermions on the Lattice
            \label{SUBSECTION_FERMIONS_FREE}}
   For simplicity let us first look at free staggered fermions on
a $d$-dimensional lattice $\L_{a/2}=(\frac{1}{2} a\bbbz)^d$ of lattice
spacing $a/2$.
 Only translations
by integer multiples $\vec{n}a$ of $a$
are regarded as true lattice translations,
i.e. survivors of translations in the continuum. These true translations
map the sublattice $\L_a = (a\bbbz)^d $ of lattice spacing $a$ into
itself.

We imagine that the lattice $\L_{a/2}$  is
divided into hypercubes of sidelength $a$ containing $2^d$ sites each.
 Sites within one hypercube are distinguished by their ``pseudoflavour''
$H$. $H$ is specified by a set
of distinct indices $\mu_i$
\be
       H = \left\{ \mu_1
\mu_2\dots \mu_h, 0\leq h < d,
                   \mu_1<\mu_2<\dots<\mu_h \right\} \es .
\label{FERMIONS_PSEUDOFLAVOR}
\ee
Sites $x_H$ in $\L_{a/2}$
with the same pseudoflavour $H$ are of the form
$$
   x_H \equiv \wh{x} +\eh e_H,\quad e_H =\sum_{\mu\in H}e_\mu,
   \quad \wh{x}\in\L_a
$$
where $e_\mu$ is the lattice vector of length $a$
in $\mu$-direction in $\L_a$. The sites $x_H$
form a sublattice $\L_a^H $ of $\L_{a/2}$
(see Fig.~\ref{FERMIONS_PSEUDOFLAVOR_1}),
viz. $\L_a^H = \L_a+ \eh e_H$.
\begin{figure}
\vspace{9.5 cm}
\caption{Part of a pseudoflavour lattice
 {$ \Lambda^H$ for $d=2$, $H \in \{ \emptyset , \{ 1 \} ,
   \{ 2 \} , \{ 1 ,  2 \} \}$},
   embedded into $\Lambda = \bbbz   \times \bbbz$.}
\label{FERMIONS_PSEUDOFLAVOR_1}
\end{figure}

Pseudoflavour
  is a combination of spin and flavour in the sense that there is
an expansion
\be
     \sum_H \chi(x_H) \ dx^H = \sum_{\a,f} \Phi_\a^{\skipp{1} f}(x)
                                        \  Z_\a^{\skipp{1} f} \es .
\label{FERMIONS_DEFINITION_STAGGERED}
\ee
    which exhibits the field $\chi(x_H)$ at site $x_H$ in a hypercube
    as a linear combination of field components
    $ \Phi_\a^{\skipp{1} f}$ with spinor index $\a$ and flavour
    index $f$ ($ Z = (Z_\a^{\skipp{1} f})$ is a matrix of differential
    forms). Each index takes values $1 , \dots , 2^{[d/2]}$.
    For details, see \cite{FERMIONS_KMS}.
\subsection*{Blocking Consistent with the Symmetry
            \label{SUBSECTION_FERMIONS_BLOCKING}}
   We distinguish between two kinds of blockings: from
  the continuum to some lattice, and from one
  lattice to a coarser lattice.
  We wish to block in such a way that as much is preserved of internal
  and space time symmetries as is possible.

\subsubsection*{Blocking from the Continuum to the Lattice.}
  Given Fermi fields $\Psi_\alpha^f(z)$ on the continuum with
  $2^{[d/2]}$ flavours $f$,
  let $\chi^H(z)$ be the linear combination of field
  components $\Psi_\alpha^f$ of given pseudoflavour.
  The staggered blockspin on the lattice $\L_{a/2} $ has the form
\be
  \Xi(x) = \int_{z} C^H(x,z)\ \chi^H(z)
  \quad\mbox{ if } x\in \L^H_a \subset \L_{a/2} \es
  \enspace .
\label{FERMIONS_STAGGERED_BLOCKSPIN}
\ee
Here $\int_z $ is an integral over continuous space.

   In the absence of gauge fields the kernel of the
averaging operator is
given by
\be
   (\frac{a}{2})^d\
    C^H(x,z) =  \left \{ \begin{array}{ll}
    1 & \quad\mbox{if $x\in\L^H_a,
                  \max\ \left|x^\mu-z^\mu\right|\leq \frac{a}{2}$}\\
    0 & \quad\mbox{otherwise}
    \enspace  .         \end{array}
               \right.
\label{FERMIONS_AVERAGE_NO_GAUGE}
\ee
i.e. one averages over a cube of sidelength $a$ with center $x$.
Note that the cubes around block lattice sites
                           $x,y$ may overlap if $x, y$ have different
pseudoflavour (see below). It can be verified that this choice of
blockspin leads to good locality properties of the effective free action
 on the lattice  \cite{Mai}.

  We proceed with considerations on the staggered fermion
symmetry group.
\subsubsection*{Residual Symmetry on the Lattice.}
  For bosons, any flavour symmetry $  G $ in the continuum
  is preserved on the lattice;
  continuous space time translations get broken to
  lattice translations and similarly for Euclidean rotations.

  For staggered fermions, the situation is
  different \cite{FERMIONS_SYMMETRY}:
  blocking as described above from the continuum, the total symmetry
  ${\cal G}= G  \times {\cal T}  $
  (e.g. flavor group $ G  = SU(2)$ for $d=4$, translation group
                                              ${\cal T}$)
   gets broken to a subgroup ${\cal G}_a$ on a lattice of lattice
  spacing $a/2$. ${\cal G}_a$ is generated by
  translations $e_\mu$ by integer multiples of $a$ and by
   certain combinations $\varepsilon d^H$ of flavour transformations
  ($\varepsilon = \pm 1$),
  and translations by $e_H/2$.
  The action $T_a$ of the lattice symmetry group on a
  staggered field on the lattice $\L_{a/2}$  reads as follows
\be
     (T_a(\varepsilon d^K) \chi)(y)
          = \varepsilon \ \check{\rho}_{H,K}
              \  \chi(y + \frac{1}{2}e_K) \\
     \quad \mbox{if} \  y \in \Lambda_a^H     \es .
\label{FERMIONS_SYMMETRY_OPERATION}
\ee
  The symbols
      $\check{\rho}_{H,K}$ denote sign factors as
  in the Clifford product of forms $\vee$, defined by
$$ dx^H \vee dx^K = dx^\mu \wedge dx^\nu + \delta^{\mu\nu}\es
   \quad\mbox{ and }\quad
  dx^H \vee dx^K = \check{\rho}_{H,K} dx^{H\triangle K}  \es ,
$$
with
the symmetric difference
   $ H\triangle K = (H\cup K) \backslash (H\cap K)$
of sets $H, K$ of pseudoflavour indices.

  This agrees with the known symmetry group of standard staggered
  fermion actions \cite{FERMIONS_STAGGER}.
 Note that translations by $e_H$ are true translations.
\subsubsection*{Fine-to-Coarse Blocking.}
  The fine-to-coarse blocking step maps fields
  which live on the fine lattice
  $\L_{a/2}$ onto fields living on
                  the coarser lattice $\L_{L_b a/2}$ of lattice spacing
  $L_ba/2$, with an integer scale factor $L_b > 1$.
  It is natural to demand that the symmetry group $\CG_{L_b a}$
  on the coarse lattice
  should be a subgroup of the symmetry group $\CG_a$ on the
finer lattice
\be
 {\cal G}_{L_b a} \subset {\cal G}_a \es .
\label{FERMIONS_GROUP_CONDITION}
\ee
  In addition to the true translations by
$e_H^{\p }= L_b e_H$ there will
be combinations
\footnote{A prime ($\p$) refers to the coarser lattice.}
$$
  \varepsilon d_H^\p \in {\cal G}_{L_b a}
$$
of flavor transformations and translations by $\frac{1}{2}e_H^{\p }$.

 The symmetry operations
$d^\p_H$ must be a combination of some $d_K$ and a true translation $e$.
 {}From~\eq{FERMIONS_SYMMETRY_OPERATION}, it can be seen that
 $d_K^\p$ contains an admixture of translation by $-e_K^\p/2$. Since
 $d_K$ contains an admixture of a translation by $\frac{1}{2}e_K$, we
 must have
\ba
    -\eh     e_H^\p &=& -\eh     e_K +    e \quad
     \mbox{with }     \quad      e =\sum_\mu      n_\mu      e_\mu
\es .
\label{FERMIONS_ODD_FACTOR}
\ea
Since $e_H^{\p }= L_b e_H$,
the only way how this can happen is that $H=K$
and $e= \frac{1}{2}(L_b-1)e_H $ =
{\it true translation}. This requires
that $L_b$ is odd,
   hence $L_b=3, 5,\dots$ but {\it not} $L_b=2$ !

   This result might have been expected, since
       only for odd $L_b$ the pseudoflavour (and parity) on the
   fine lattice $\L_{a/2}$ and on the block lattice $\L_{L_ba/2}$
   match   nicely when we
                    regard $\L_{L_ba/2}$ as sublattice of $\L_{a/2}$
   (see Fig.~\ref{FERMIONS_OVERLAP} below).

   For a true symmetry it should not matter in which order we apply
   the blocking step and the symmetry operations: the blocking map $C$
   should commute with the action  $T$ of the symmetry under
   ${\cal G}_{L_ba}\subset{\cal G}_a$, viz.
\be
   T_{La}(d^\p_H)\ C(x,z) = C(x,z)\ T_a(d_H)\
   T_a(\eh \left(L_b-1\right) e_H)\es .
\label{FERMIONS_COMMUTATION_CONDITION}
\ee
 This is fulfilled if $C$ is translation invariant so that
\be
   C(x+e_H^\p, z+L_b e_H) = C(x,z)
\label{FERMIONS_TRANSLATION_INVARIANCE}
\ee
 and if in addition $C$ has the support property
\be
   C(x,z) = 0
   \quad\mbox{ unless $x,z$ carry the same pseudoflavour.}
\label{FERMIONS_C_SUPPORT}
\ee
In words, the support  property states the following. If %
$y\in\L^H_{L_ba/2}$ %
then the (staggered) blockspin $\Xi(y)$ is a weighted %
sum of fields $\chi(z)$ at sites with the same pseudoflavour $H$. %
 Blockings with such a property were considered for 2-dimensional %
$U(1)$ gauge fields by Ben Av et al. \cite{BenBraSol}. %
 However, they chose a scale factor $L_b=2$.
\subsection*{Generalization to Nontrivial Gauge Fields
            \label{SUBSECTION_FERMIONS_GAUGE}}
We now admit dynamical gauge fields in our staggered fermion
theory; the averaging kernel $C$ will then have to depend on the
gauge field for reasons of covariance as
in~\eq{WHAT_WITH_GAUGE_AND_FERMIONS}. Our interest will gradually
shift towards a practical implementation of the
renormalization group picture outlined in
Sect.~\ref{SECTION_RG_PICTURE}, known as
the {\it multigrid approach}
to numerical computations in lattice field theories~\footnote{
Those readers who are not familiar with multigrid methods
may first want to read
Sect.~\ref{SECTION_PROPAGATOR}; a
more detailed mathematical account can be found in
   \cite{FERMIONS_HACKBUSCH}.}.

To specify a blocking prescription, we retain
blocks as introduced in Sect.~\ref{SECTION_BOSONS}:
  the solution of the gauge covariant eigenvalue equation
\be
 -\Lap_{b.c.} C^\ast(z,x) = \lambda_0(x) C^\ast(z,x)
\label{FERMIONS_EVE}
\ee
  yields a gauge covariant kernel $C$.
Results reported below show that~\eq{FERMIONS_EVE} produces a good
blockspin when $\Delta_{b.c.}$ is the Laplace operator with
appropriate boundary conditions (basically Neumann b.c.).
The covariant Laplace operator
connects only lattice sites of the same pseudoflavour;
therefore~\eq{FERMIONS_EVE} is consistent with the
condition~\eq{FERMIONS_C_SUPPORT} above.

 In a multigrid approach, one needs interpolating kernels $\A $ from
 coarse to fine lattices in addition to the averaging kernels $C$.
 The kernels $\A $ should be smooth.
                               In the present context,
  there are a priori two possible definitions of smoothness.
\subsubsection*{1.\ ``Laplacian'' smoothness.}
$\chi $ possesses Laplacian smoothness if
$$ (\nabla_{\mu }\chi , \nabla_{\mu }\chi )\equiv (\chi , -\Lap \chi )
$$
is small, with
the {\it covariant Laplacian $\Lap$}:
for our model of staggered fermions this is expected to work if
the field strength tensor
$F_{\mu\nu}$, conveniently defined on the lattice \cite{MacDLGT} by
\begin{equation}
    F_{\mu\nu}(z)=   U(z+\frac{1}{2}e_\mu, \nu)\ U(z,\mu)
                  -  U(z+\frac{1}{2}e_\nu, \mu)\ U(z,\nu) \  .
\label{FERMIONS_EFFMUNU}
\end{equation}
is  small. This is the standard
definition of smoothness.

The problem is that
there are no smooth functions in this sense if the
gauge fields are disordered, because $\-\Lap $ is strictly positive in
this case.

\subsubsection*{2.\ ``Diracian'' smoothness.}
$\chi $ possesses Diracian smoothness if
$$ (\Dirac \chi , \Dirac \chi )\equiv (\chi , -\Dirac^2 \chi ) $$
is small.
Since
\be
                  -\Dirac^2 = -\Lap + \sigma_{\mu\nu}F_{\mu\nu} \ ,
\label{FERMIONS_DIRACIAN}
\ee
 sites of different pseudoflavour are connected.

When one wants to compute propagators of fermions by use of the
multigrid approach, then one should in principle use interpolating
kernels $\A $ which possess Diracian smoothness. Unfortunately this
is impractical for reasons of storage space. Laplacian smooth
kernels $\A $ are much more practical. Possible compromises are
discussed in \cite{FERMIONS_KMS}.

In Sect.~\ref{SECTION_NEURAL} the problem of smoothness of the
interpolation kernel $\A$ (see Sect.~\ref{SECTION_BOSONS})
will be readdressed using optimization wisdom borrowed from
neural computation.
\subsection*{Special Choice which yields a Good Blockspin
            \label{SUBSECTION_FERMIONS_GOOD}}
In the language of Sect.~\ref{SECTION_BOSONS},
we know that we have made a good choice of blockspin if
the effective action remains {\it local} after arbitrarily many or
arbitrarily big blocking steps.
The averaging kernel $C(x,z)$
                           which is nonvanishing only
if $z$ and $x$ have the same pseudoflavour, and which satisfies the
Laplace equation~\eq{FERMIONS_C_SUPPORT} with
Neumann boundary conditions
  on block
boundaries,                                 has proven to be good
in that sense.
   This result extends to blocking from continuum action with
   $2^{[d/2]}$ flavours to staggered fermion field on the lattice.
               Details will be given in the next section.
Fig.~\ref{FERMIONS_OVERLAP} visualizes the overlapping of blocks
which do not share sites of the same pseudoflavour as the block centers.
\begin{figure}
\vspace{10.0 cm}
\caption{The good choice of blocks to average over. $C(z,x)$ is only
   nonvanishing
   if site $z$ has the same pseudoflavour (symbol) as the block-center
   $\wh{x}$ (encircled symbol). Therefore the seemingly
   overlapping blocks
   (square) have actually no sites in common.}
\label{FERMIONS_OVERLAP}
\end{figure}
 In numerical investigations
\cite{Vin}
 non-overlapping blocks were used as well. But then
     one does not retain the successful pure gauge limit.

A successful block spin requires $2^{[d/2]}$ flavors.
There is no consistent way of putting a single chiral particle
on the lattice (this is the statement of the theorem by
Nielsen and Ninomiya~\cite{FERMIONS_NIELSEN_NINOMIYA}):
Therefore there is no good blockspin either for a single chiral field,
as we had already remarked in Sect.~\ref{SECTION_WHAT}.
\section{Efficient Computation of Gauge Covariant Propagators
         \label{SECTION_PROPAGATOR}}
 A quenched simulation of lattice QCD requires the evaluation of
 products of Dirac propagators $S(U)$.
 $S$ depends on the actual gauge field configuration $U$.
 In a Monte Carlo simulation of full QCD, one needs also the fermion
 determinant
 $$
    {\rm det}^{-1/2}S(U) =
       \N^{-1} \int \D\phi\ \exp\left[ -(\phi ,S(U)\phi )\right]\es .
 $$
 Since for every new gauge field configuration $U$, $S(U)$ needs to be
 recomputed or updated, it is important to have an efficient
 method to do so.
 The conjugate gradient (CG) or the minimal residual algorithm
 is state of the art.
 Great hopes to do better are attached to multigrid (MG) methods
 \cite{BenBraSol,Vin,BMMR,BRV,HSVLAT90,BERV,Vya,KalPL}.

\subsection*{Deterministic Multigrid Methods
            \label{SUBSECTION_PROPAGATOR_DETERMINISTIC}}
For convenience, we will use the opposite convention in enumerating
the lattices of a multigrid in this section. The fundamental (i.e.
finest) lattice is denoted by $\Lambda =\Lnull$. There will be
a sequence of lattices $
   \Lnull$, $\Lone$, $\Ltwo$, $\ldots$ of increasing lattice
 spacing $a_i$, viz.\ $a_{i+1} = L_b a_i$ with $a_0 = a$.
 We wish to solve an inhomogeneous linear equation
 \begin{equation}
  D_0 \chi_0 = f_0
 \label{8.1}
 \end{equation}
 on the fundamental lattice $\Lambda = \Lnull$, for given $f_0$.
 In our case, $D_0 = -\Dirac^2 + m^2$.
 After some relaxation sweeps on $\Lnull$ one gets an approximation
 $\tilde{\chi}_0$ to $\chi_0$ which differs from the exact solution by
 an error $e_0 = \chi_0 - \tilde{\chi}_0$.
 The fundamental idea of the MG to the solution of partial
 differential (or difference) equations
 \cite{FERMIONS_HACKBUSCH,MG} is that the
 error $e_0$ should become smooth very fast, although it may become
 small very slowly because of critical slowing down (CSD).
 The error satisfies the equation
 \begin{equation}
  D_0 e_0 = r_0
 \label{8.2}
 \end{equation}
 which involves the residual $r_0 = f_0 - D_0 \tilde{\chi}_0$.
 If $e_0$ is smooth, it is determined to a very good accuracy by a
 function $e_1$ on the next coarser lattice $\Lone$, and can be
 represented in the form
 \begin{equation}
  e_0 = \A e_1
 \label{8.3}
 \end{equation}
 with an interpolation map $\A$ which should be so chosen that it maps
 functions on $\Lone$ into smooth functions on $\Lnull$.
 Conversely, $e_1$ can be obtained from $e_0$ with the help of an
 averaging map $C$ which satisfies
 \begin{equation}
  C \A = \bbbone \es .
 \label{8.4}
 \end{equation}
 It follows that $e_1 = C e_0$.
 Inserting~\eq{8.3} into~\eq{8.2} and acting on the result
 with $C$, we see that $e_1$ will satisfy the equation
 \begin{equation}
  D_1 e_1 = r_1
 \label{8.5}
 \end{equation}
 with
 \begin{equation}
  D_1 = C D_0 \A \ \ \ , \ \ \ r_1 = C r_0 \es .
 \nn
 \end{equation}
 The problem has been reduced to an equation on the coarser lattice.
 If there is still too much CSD at this level,
 one may repeat the procedure, going to coarser and coarser lattices.

 Given $\A$, a possible choice of $C$ which satisfies~\eq{8.4} is
 $C = \left( \A^* \A \right)^{-1} \A^*$.
The kernel $\A (z,x)$ should be a smooth function of $z$.
 The appropriate notion of smoothness depends on the dynamics,
 i.\ e.\ on $D_0$, in general.
 {\em Smooth means little contributions from eigenfunctions to high
 eigenvalues of $D_0$.}
 This insight is confirmed by results reported below.
 The point is important in systems in gauge fields and for other
 disordered systems.

\subsection*{Gauge Covariance
            \label{SUBSECTION_PROPAGATOR_COVARIANCE}}
 In a gauge theory, kernels $\A$ and $C$ should be chosen in a gauge
 covariant fashion;
 see the discussion in Sect.~\ref{SECTION_BOSONS}.
 Under gauge transformations on $\Lnull$ they transform according to
 \begin{eqnarray}
      \A (z,x) &\mapsto& g(z) \A (z,x) g(\wh{x})^{-1}\es , \nonumber \\
        C (x,z) &\mapsto& g(\wh{x}) C (x,z) g(z)^{-1} \es .
 \label{covofkernels}
 \end{eqnarray}
 This is consistent with $C\A = \bbbone$.

 The most general expression of a kernel $\A$ with this covariance
 property is a weighted sum of parallel transporters $U({\cal C})$
 along paths ${\cal C}$ from $\wh{x}$ to $z$, e.\ g.\
 \begin{equation}
    \A (z,x) = \sum_{{\cal C}:\,\hat{x}\,\mapsto z}
 \varrho({\cal C})\ U({\cal C}) \es ,
 \label{sumofpaths}
 \end{equation}
 where $\varrho({\cal C})$ are numbers.
 And analogously for $C$.
 We will prefer not to specify the weights $\varrho$ explicitly,
 but to determine $\A$, $C$ as solutions of covariant equations.

\subsection*{Ground-State Projection Multigrid
            \label{SUBSECTION_PROPAGATOR_GSPMG}}
 In the ground-state projection MG method
 the averaging operator $C$ from a grid to the next coarser
 grid is a projector on the ground-state of a local Hamiltonian.
 The adjoint of $C$ satisfies a gauge covariant eigenvalue equation
 (cf.~\eq{EVC},~\eq{FERMIONS_EVE}),
 \begin{equation}
   (- \Delta_{\subs{b.c.}} C^*)(z,x) = \epsilon_0(x) C^* (z,x)\es .
 \label{EVEQC}
 \end{equation}
 Remember that the kernel $C^* (z,x)$ which solves~\eq{EVEQC}
 is an $N_{\subs{c}} \times N_{\subs{c}}$ matrix which is in general
 not an element of the gauge group.
 ($N_{\subs{c}}$ is the number of colours.)

 For bosons one may choose Neumann boundary conditions on block
 boundaries,  $- \Delta_{\subs{b.c.}}  =  -\Delta_{N,x}$, as
 discussed in Sect.~\ref{SECTION_BOSONS},~\eq{EVC}.
 The lowest eigenvalue $\epsilon_0(x)$ of $-\Delta_{\subs{b.c.}}$
 is a gauge invariant quantity which is a measure of disorder.
 A normalization condition $C C^* = \bbbone$ may be imposed,
 and the covariance condition $C^*(\hat{x},x) > 0$ (as a matrix) makes
 the definition of $C$ unique.
 Kernels defined by~\eq{EVEQC} and  $C^*(\hat{x},x) > 0$
 enjoy gauge covariance~\eq{covofkernels}.

 For staggered fermions,
 the Laplacian should be regarded as living on the sublattices
 which consist of sites with the same pseudoflavor. It
               connects only sites with the same pseudoflavor.
 Neumann boundary conditions on the Laplacian on the sublattice of
 sites $z$ with the same pseudoflavor as $x$ must be supplemented by
 the requirement that $C(x,z)=0$ unless $z$ and $x$ have the same
 pseudoflavor. This extra requirement could be regarded as a consequence
 of supplementary boundary conditions.

 There exists an efficient algorithm for computing $C$
 \cite{KalNP}.
 The programs are fully vectorized.
 In 4-dimensional $SU(2)$ gauge fields,
 computation of $C$ on the whole lattice costs CPU time of the
 order of one heatbath sweep for the gauge field.
 Thus, it is not too expensive to solve~\eq{EVEQC}, and it is not
 necessary to sacrifice gauge covariance by working with gauge field
 independent kernels in gauge-fixed $U$-configurations.
 (Gauge fixing costs also CPU time.)

\subsection*{Interpolation Kernel $\A$
            \label{SUBSECTION_PROPAGATOR_INTERPOLATION}}
 Given the averaging kernel $C$, there exists an ideal choice
 of the interpolation kernel $\A$.
 It is determined as follows.
 For every function (``blockspin'') $\Phi$ on $\Lone$, $\phi = \A \Phi$
 minimizes the action $\H = \left( \phi , D_0\phi \right)$
 subject to the constraint $C \phi = \Phi$.
 Recalling the definition of smoothness,
   this can be rephrased:
 $\phi = \A \Phi$ is the smoothest possible function with
 prescribed block average $\Phi = C \phi$. It follows that $\A $
 satisfies \eq{CondAC} below.
 With this choice of $\A$,
 the action $\H$ completely decouples into a sum of actions for
 the different MG layers. Moreover, $D_1=(CC^{\ast})\A^{\ast} D_0\A $;
 this is selfadjoint for our $C$-kernels which satisfy
 $CC^{\ast }= \bbbone$.

 A good ``choice of blockspin'', i.e. of $C$, is characterized
 by the fact that the ideal kernel $\A (z,x)$ ($z \in \Lnull$,
 $x \in \Lone$) associated with it has good locality properties.
 This means that $\A (z,x)$ is big for $z \in x$, and decays
 exponentially in $| z - \hat{x} |$ with decay length 1 block lattice
 spacing $a_1$ (cf.\ Sect.~\ref{SECTION_EFFECTIVE}).
 Numerical computations of the ideal kernel $\A$ were done
 in quenched 4-dimensional $SU(2)$ gauge fields at various values
 of $\beta = 4/g^2$, including the case of a
 completely random gauge field ($\beta = 0$).
 (For bosons $D_0=-\Delta + m^2 $,
  while for fermions $D_0 = -\Dirac^2+m^2$ with $m^2=0$ or small).
 It was verified that for {\em any} $\beta$-value
 the definition of the averaging kernel $C$ as a solution of a
 gauge covariant Laplace eigenvalue
 equation~\eq{EVC},~\eq{FERMIONS_EVE}
 (with Neumann boundary conditions on block boundaries)
 yields a good choice of blockspin in the sense described
 above. This is true both for bosons and for staggered fermions;
        see Figs.~\ref{FigBoseA},~\ref{FigFermiA}.
 $\A$ decays exponentially over distance $a_1$ (nearly) as fast
 in the presence of gauge fields as in their absence.
 \begin{figure}
 \vspace{11.0 cm}
 \caption{Optimal interpolation kernel $\A (z,y)$ for bosons on an
          $18^4$ lattice in a quenched $SU(2)$ gauge field equilibrated
          at $\beta = 2.7$.
          ($L_b = 3$, periodic b.\ c.)
          Shown is a two-dimensional cut through the block center $y=0$,
          $z_3$ and $z_4$ are fixed.
          The vertical axis gives the trace norm of $\A (z,y)$.
          (Remember that $\A (z,y) \in \bbbr \cdot SU(2)$.)
          \label{FigBoseA}}
 \end{figure}

 Because of the exponential decay of the optimal kernel $\A$,
 one may try to approximate it by a simpler choice which fulfills
 $\A(z,x)=0$ for $z\not\in$ a certain neighborhood of block $x$.
 This neighborhood might be taken to be only the block $x$ itself.
 The coarse grid operator would then be defined through
 $D_1 = \A^* D_0 \A$.

 It follows from the optimization criteria discussed below that $\A$
 should have the property that the lowest eigenmode $\phi_0$ of
 $D_0$ should admit a representation of the form $\phi_0 = \A \Phi$
 with a suitable $\Phi$ to a very good accuracy.
 The proposal of \cite{KalNP} to define $\A$ through the eigenvalue
 equation~\eq{EVEQC} with $-\Delta_{\subs{b.c.}}$ taken to be a
 local Hamiltonian with Dirichlet instead of Neumann boundary
 conditions does not meet this requirement.

 {\em For periodic boundary conditions} the Galerkin choice (also
 called variational coarsening) $\A = C^*$ is better.
 In this case $D_1$ can be defined as $D_1 = \delta^{-1} C D_0 C^*$
 \cite{KalPL}.
 The fluctuating length of its matrix elements can be adjusted to
 fluctuate around the ``right'' value (of the exact $D_1 (x,y)$) by
 tuning the real parameter $\delta$.

\subsection*{Numerical Results for Propagators Computed by MG
            \label{SUBSECTION_PROPAGATOR_RESULTS}}
 Ground-state projection MG computations for propagators were done
 in 2-$d$ $U(1)$ \cite{BRV,HSVLAT90} and (approximate ground-state
 projections) in 2-$d$ and 4-$d$ $U(1)$ and 2-$d$ $SU(2)$ gauge fields
 \cite{BRV,BERV,Vin}.
 The first ground-state projection MG computation of gauge covariant
 propagators in 4-$d$ non-Abelian gauge fields (for gauge group
 $SU(2)$) was presented in Ref.~\cite{KalPL}.
 Besides ground-state projection there is another approach for choosing
 the averaging kernel $C$.
 This approach is called ``parallel transported MG''.
 It is based on the representation~\eq{sumofpaths} and is studied
 in Refs.~\cite{BenBraSol}.

\subsubsection*{Bosonic Propagators in 4-$d$ $SU(2)$ Gauge Fields.}
 For bosons, $D_0 = -\Delta + m^2 $.
 The MG method is of interest near criticality, i.\ e.\ for slowly
 decaying propagators.
 For nontrivial gauge fields we enforce slow decay by choosing $m^2$
 negative and very close to the negative of the lowest eigenvalue
 $-\mcr > 0$ of $-\Delta$.%
 \footnote{Relaxation times of conventional iterative algorithms
           depend only on $m^2$ and {\em not} on the lattice size
           $|\Lambda |$;
           this is contrarily to the Dirichlet problem, for instance.
           In case of periodic b.\ c.\ there is only an implicit
           (slight) dependence on $|\Lambda |$ through the value of
           $\mcr$.}
 The relaxation time $\tau$ -- defined by the {\em asymptotic}
 exponential decay of the norm of the residual or error --
 behaves like $\tau \propto (\Dm)^{-z/2}$ for small $\Dm =  m^2 - \mcr$.

 \medskip\noindent
 Results
 in 4-dimensional $SU(2)$ gauge fields equilibrated with Wilson's
 action at various values of $\beta$ \cite{KalPL}:
 \begin{itemize}
  \item The MG algorithm with the ideal interpolation kernel $\A$
                       eliminates CSD for {\em any} value
        of the gauge coupling; see Table~\ref{TableOptMG}.
        These computations were the first without CSD in non-trivial
        gauge fields.
        They prove that {\em ground-state projection is a good choice of
        $C$}, and that {\em the MG method can cope with the
        ``frustration'' (disorder) which is inherent in non-Abelian
        gauge fields}.
        However, the optimal $\A (z,y)$ is not translational invariant
        (except for $U \equiv \bbbone$) and has support on all sites
        $z$ of the fine lattice, for all sites $y$ of the block lattice.
        Therefore the use of this kernel for production runs is
        impractical, but it was important to answer questions of
        principle.
        In particular this result confirms that there exists an
        appropriate notion of smoothness in disordered cases,
        as described above.
        The ideal $\A$ is the smoothest kernel which obeys
        $C \A = \bbbone$.
  \item The most practical MG algorithm with $\A = C^*$ does not
        eliminate CSD in non-trivial gauge fields, $z=2$ remains
        (as in 1-grid relaxation), although $z=0$ in pure gauges;
        see Fig.~\ref{FigTau}.
        However, MG is able to outperform the conjugate gradient
        algorithm; see Fig.~\ref{FigConvergence}.
  \item After the publication of Ref.~\cite{KalPL}, a practical
        nonlinear MG algorithm has been developed which eliminates
        the appearance of CSD when the mass is lowered,
        some volume dependence of the correlation times still
                      remains, though.
        Details will be reported elsewhere \cite{Kalinprep}.
 \end{itemize}
 \begin{table}
 \caption{Results of the idealized MG algorithm with lexicographic SOR
          (relaxation parameter $\omega$) on $9^4$ lattices $\Lnull$.
          Kernel $\A$ is the solution of the equation
         $\left( [ - \Delta + \mcr + \kappa\,\Cstar C ] \A \right) (z,y)
          = \kappa\,\Cstar (z,y)$ for large $\kappa$.
         This expression tends to the ideal $\A$-kernel described in
         the text for large $\kappa$.
         \label{TableOptMG}}
 \begin{center}
 \begin{tabular}{|c|c|c|c|}
 \hline
 $\beta$ & $\mcr$ & optimal $\omega$ & $\tau$ for $\Dm \leq 10^{-3}$ \\
 \hline\hline
 $\infty$ & $0$          & $1.27$ & $1.6$ \\
 \hline
 $2.7$    & $-0.8210607$ & $1.38$ & $1.9$ \\
 \hline
 $2.5$    & $-0.9477085$ & $1.40$ & $1.9$ \\
 \hline
 $2.2$    & $-1.2218471$ & $1.45$ & $1.9$ \\
 \hline
 $1.8$    & $-1.7567164$ & $1.57$ & $2.5$ \\
 \hline
 $0$      & $-2.7480401$ & $1.69$ & $5.2$ \\
 \hline
 \end{tabular}
 \end{center}
 \end{table}
 \begin{figure}
 \vspace{7.5 cm}
 \caption{Computation of bosonic propagators $(-\Delta + \mcr +
          \triangle m^2)^{-1}$.
          Relaxation times $\tau$ (in comparable work units) of
         iterative algorithms on an $18^4$ lattice in a quenched $SU(2)$
         gauge field at $\beta = 2.7$  with $\mcr = -0.7554339$.
         (MG with $\A = C^*$; Jac means Jacobi relaxation, while SOR
         stands for successive over-relaxation.)
         \label{FigTau}}
 \end{figure}
 \begin{figure}
 \vspace{10.0 cm}
 \caption{Computation of bosonic propagators $(-\Delta + \mcr +
          \triangle m^2)^{-1}$.
          Convergence on an $18^4$ lattice in quenched $SU(2)$ gauge
         fields at (a) $\beta = 10.0$, and (b) $\beta = 2.7$, with
         $\mcr = -0.1533739$, resp.\ $-0.7554339$.
         (MG with $\A = C^*$.)
         \label{FigConvergence}}
 \end{figure}
\subsubsection*{Propagators for Staggered Fermions in 4-$d$
               $SU(2)$ Gauge Fields.}
 Of the two different proposals for the averaging kernel $C$ made
 in \cite{FERMIONS_KMS}, only the Laplacian choice has been
 implemented numerically yet.
 There, $-\Delta_{\subs{b.c.}}$ = $-\Delta_{N,x}$ in~\eq{EVEQC},
  where $\Delta$ is the fermionic 2-link Laplacian defined
 in~\eq{FERMIONS_DIRACIAN}.
 The optimal interpolation kernel $\A$ associated with this $C$
 has the desired falloff properties in non-trivial gauge fields;
 see Fig.~\ref{FigFermiA}.
 Thus, the Laplacian choice for $C$ defines a good blockspin
 for staggered fermions, and therefore it seems not to be necessary to
 implement the more intricate proposals mentioned in
 \cite{FERMIONS_KMS}.
 However, numerical simulations with the optimal $\A$ have not been
 performed yet.
 \begin{figure}
 \vspace{11.0 cm}
 \caption{Optimal interpolation kernel $\A (z,y)$ -- associated
          with the ``Laplacian'' choice for $C$ -- for staggered
          fermions on an $18^4$ lattice in a quenched $SU(2)$ gauge
          field equilibrated at $\beta = 2.7$.
          ($L_b = 3$, periodic b.\ c.)
          $\A$ is the solution of the equation
          $\left( [ - \Dirac^2 + \kappa\,\Cstar C ] \A \right) (z,y)
          = \kappa\,\Cstar (z,y)$ for large $\kappa$.
          Shown is a two-dimensional cut through the block center $y=0$,
          $z_3$ and $z_4$ are fixed.
          The vertical axis gives the trace norm of $\A (z,y)$.
          (Remember that $\A (z,y) \in \bbbr \cdot SU(2)$.)
          Note the support properties of $\A$, it has support on all
          even lattice sites and is vanishing on all odd sites.
          This is due to the non-vanishing field strength at finite
          $\beta$.
          In the limiting case of pure gauges the field strength term
          $\sigma_{\mu\nu} F_{\mu\nu}$ vanishes and the support of
          $\A$ is reduced to $1/16$-th of $|\Lambda |$.
          Then we have 16 bosonic $\A$-kernels on $9^4$ sublattices.
          \label{FigFermiA}}
 \end{figure}

 MG with variational coarsening $\A = C^*$ works, but up to now
 we had to restrict ourselves to relatively small lattices (up to
 $18^4$) where this MG algorithm is not competitive with conjugate
 gradient.
 The authors of Refs.~\cite{BenBraSol} report that their algorithm
 performs considerably better when they re-unitarize the blocked
 gauge field (matrix elements of the effective Dirac operator)
 such that it is again an element of the gauge group.
 However, our results indicate that the difference between using
 a unitary or a dielectric gauge field \cite{MacDLGT} on coarser
 layers can be compensated by adjusting the real parameter $\delta$
 in $D_1$.
 By tuning $\delta$, one can re-unitarize the blocked gauge field
 ``on the average''.

 We are optimistic that the implementation of the nonlinear
 MG algorithm found for bosons (see above) also gives a practical
 and competitive MG method for fermions.
 This implementation is currently investigated.
 For details refer to \cite{Kalinprep}.
\section{Neural op\-ti\-mi\-za\-tion of de\-ter\-mi\-nis\-tic
         mul\-ti\-grids \label{SECTION_NEURAL}}
The results of Sect.~\ref{SECTION_PROPAGATOR}
                         confirm that the multigrid approach is able
to handle inhomogeneous wave equations in disordered media, e.g. in
the presence of gauge fields, but the appropriate notion of smoothness
depends on the dynamics, i.e. on the operator $D_0$ in the equation
which one wishes to solve. The interpolation kernels $\A $
                                                     should be smooth.
Smooth means little contributions from eigenfunctions to high
eigenvalues of $D_0$.

This means that the appropriate
                    kernels $\A$ are dynamically determined and need to
be found before or together with the solution of the equation.
(The averaging kernel $C$ is often related to $\A $.)
The question is how.
                                           There are interesting
potential applications other than gauge theories
\footnote{                                        For instance,
one may admit link variables $U(z,w)$ in the covariant Laplacian
$\Delta$ which take values $0$ or $1$. Links $(z,w)$ with $U(z,w)=0$
are considered deleted from the lattice. If deleted links are
distributed at random with suitable probability, on obtains a fractal
lattice. Solutions of the equation $(-\Delta +M^2)\chi_0 = f_0$
admit a random walk representation on this lattice.}.
Therefore a general strategy is desired.
\subsection*{Criteria for optimality
            \label{SUBSECTION_CRITERIA}}
One may write down a cost functional whose minimum yields the optimal
kernels, i.e. the fastest possible convergence of the multigrid
iteration.

For definitenes and simplicity, think of a twogrid. Assume that
updating consists of standard damped
                              relaxation sweeps on the fundamental
lattice to smoothen the errors, followed by an updating step
$\chi_0 \mapsto \chi_0 - \A e_1 $ which involves the exact solution
$e_1$ of~\eq{8.5}  on the coarse grid $\Lambda^1$,
                   with some $D_1$ which is to be determined together
with $\A $ and $C$.

Convergence speed is estimated by the norm of the iteration matrix
$\rho $. The iteration matrix is the product $\rho = \rho_0\rho_1$
of the iteration matrix $\rho_0$ for the smoothing, and the iteration
matrix $\rho_1$ for the coarse grid updating. The norm
$$ \|\rho \| \leq \| D_0 \rho_0\| \cdot \|D_0^{-1}\rho_1\| \es . $$
Only the second factor depends on $\A $, $C$ and $D_1$.
                                                        Inclusion of
the operators $D_0^{\pm 1}$ in the factors reflects the philosophy
that smoothing suppresses the high frequency modes of the error
\cite{FERMIONS_HACKBUSCH}. Choosing the trace norm, one is lead to
determine $\A $, $C$ and $D_1$ from the minimality condition
for $E_1= \|D_0^{-1}\rho_1\|^2 $ ,
\ba
  \min &=& E_1\ = \int \int_{z,w\in \L^0}  |\Gamma (z,w)|^2 \\
  \G(z,w)&=& v(z,w)-\int\int_{x,y\in\L^1}
  \A(z,x)\ \tilde{u}(x,y)\ C(y,w) \es  \\
  v &=& D_0^{-1} \ \ \mbox{and } \ \ \tilde{u}=D_1^{-1} \es
\ea
The minimality condition leads to conditions on kernels $\A $, $C $
and on $D_1$. Suppose first that no
constraints of practicality
are imposed on the kernels. We assume that $D_0$ is selfadjoint.
Optimizing $\A $, for given $C$ and ${\tilde u}=D_1^{-1}$, one finds the
condition
$ \Gamma C^{\ast }  = 0$. That is,
\be D_0\A = C^{\ast }u^{-1} \label{CondAC} \ee
with $u= {\tilde u} CC^{\ast }$.
This reproduces the familiar condition for an ideal $\A $-kernel.
Variation of the cost functional with respect to $C$ yields
$\A^{\ast }\Gamma = 0.$ This is equivalent to a
      second condition which is obtained from~\eq{CondAC} by
interchange of the role of $\A $ and $C^{\ast }$ and substituting
$D_1^{\ast }$ for $D_1$. Variation with respect to $D_1$ yields no new
condition. These conditions leave  much freedom.
\footnote{For instance, one may demand that $C=(\A \A^{\ast })^{-1}
\A^{\ast} $ or, equivalently, $\A = C^{\ast }(CC^{\ast })^{-1}$. Then
$C\A = \bbbone $, the second condition follows from the first, and
$\Gamma $ agrees with the fluctuation field propagator \eq{GammaExp}
with $D_1=-CC^{\ast }\Delta_{eff}$ if $D_0=-\Delta $. }

But practicality requires that one imposes restrictions of the form
$\A (z,x)=0$ except when $z$ is sufficiently close to $x$, and
similarly for $C$ and $D_1$.
(We will regard $\L^1 $ as a sublattice of $\L^0$.) The proposal is
to find the optimal kernels which satisfy these constraints
                            by minimizing $E_1$.

The idea was tested on a baby model, the 1-dimensional Poisson
equation
\be
          (-\Delta + M^2 )\ \chi_0 = f_0
\label{NEURAL_POISSON}
\ee
 on a chain $\L^0$ of
lattice spacing $1$, with very small $M$.
$\L^1 $ consisted of the even sites in $\L^0$. As a condition of
practicality it was demanded that
$$ C(x,z)= 0 = \A (z,x) \quad \mbox{ unless } \quad |z-x|\leq 1 \es .$$
Optimization via gradient descent
             yielded the expected result, in the limit of small $M$,
\ba
    C(x,x)&=& 1\quad,\quad
                  C(x, x\pm 1) = \frac 12  \quad , \quad
                  D_1=2(-\Delta_1+4M^2)\\
    \A(z,x)&=&C(x,z) \quad\quad (x\in \L^1 , z\in \L^0)   \es ,
\ea
modulo normalization conventions. $\Delta_1$ stands for the
Laplacian on $\L^{1}$.
\subsection*{Neural net approach
            \label{SUBSECTION_NEURAL}}
It is interesting to rephrase the problem in the language of neural
nets \cite{Palmer}.

We regard the layers of the multigrid as layers of a perceptron. The
sites of the lattices are called nodes in this context. In the twogrid
case there will be an input layer with nodes $z\in \L^0$, a hidden
layer with nodes $x\in \L^1 $ which are sites of the coarse grid, and
an output layer with nodes $w\in \L^0$. Nodes $z$ of the input layer
are connected to nodes $x$ of the hidden layer by ``neurons'' with
connection strength $C(x,z)$, so that node $x$ receives as input
                                 a linear superposition of the
output of the input nodes $z$. Nodes $x$ of the hidden
layer are connected to nodes $w$ of the output layer by ``neurons'' of
connection strength $\A (w,x)$ (see Fig.~\ref{NEURAL_TWOGRID}).
\begin{figure}
\vspace{5.0 cm}
\caption{Neural net from a twogrid $(\L^0,\L^1)$ in one dimension. It
 has
 one input layer $\L^0$, one output layer $\L^0$ and one layer $\L^1$
 of hidden nodes.}
\label{NEURAL_TWOGRID}
\end{figure}

The network receives as input an approximation $\xi (z)$ to the
solution of the
equation~\eq{8.1}
which we wish to solve, and performs an iteration step
in order to obtain as output an improved approximation $O(z)$.
The desired
output is the exact solution $\zeta (z)=D_0^{-1}f_0$. If the iteration
step involves the exact solution of the coarse grid equation~\eq{8.5}
then the network is not strictly feed-forward.
\footnote{A neural network is called ``feed-forward'' if the nodes in a
layer receive their only input from nodes on preceding layers
\cite{Palmer}.}
                                               This is
                                              because the hidden layer
needs to do computations more general than the usual computation of
some nonlinear
 function at individual nodes in order to solve the coarse
grid equation. This could be remedied if one envisages iterative
solution of the coarse grid equation~\eq{8.5}.
                                      We return to this below.

The network is supposed to learn to compute an output $O$ which is as
close as possible to the target $\zeta$. The learning process
consists in adjusting the connection
strengths $C(x,z)$ and $\A (z,x)$. One envisages that a training set
$(\xi^{\mu }, \zeta^{\mu })$ is presented to the network to teach it.
It involves a sequence of inputs $\xi^{\mu }$. The network has its
output $O^{\mu }$
 compared to the desired answer $\zeta^{\mu }$ and receives
feedback about the error which is then used to adjust the connection
strengths according to some learning rule. Application of
typical learning rules amounts to iterative minimization of some
cost functional $E$. The standard cost functional is
 $$ E = \sum_{\mu } \sum_{z} |O^{\mu }(z)-\zeta^{\mu }(z)|^2 \es .$$
In our case, the output is a linear function of the input and of $f_0$
which has the exact solution $D_0^{-1}f_0$ as a fixed point. Therefore
 $$ O = \rho\ \xi  + (1-\rho) D_0^{-1} f_0 \ , $$
where $\rho $ is the iteration matrix. The target $\zeta^{\mu}
= D_0^{-1}f_0$ is independent of the input. The cost functional becomes
$$
   E=\sum_{\mu } \left\|\rho
                  \left(\xi - D_0^{-1}f_0\right)
                                                \right\|^2
$$
where $\|f\|^2=\langle f,f\rangle=\sum_z |f(z)|^2 $.
Since the equation is linear, we
may imagine without loss of generality that $f_0$ is arbitrarily small.
Assuming the training input is a complete orthonormal set $\xi^{\mu }$
of functions on $\L^0$ in the limit of small $f_0$,
                        the cost functional becomes
$$ E = \|\rho \|^2 \equiv \tr( \rho \rho^{\ast}) \es .$$
This is the same cost functional as considered before.

We could restrict our attention
to the coarse grid updating whose updating matrix is
$\rho_1$.                                       In this case
                                                one wants mainly
to suppress the low frequency modes of the error. This motivates us to
propose a training set with $\xi^{\mu }$ such that $D_0 \xi^{\mu } $ are
 orthonormal. As a result one finds that $E_1$ should be minimized,
where $E_1= \|D_0^{-1}\rho_1\|^2 $ as before. Minimization
of $E_1$ leads to less practical learning rules than $E$
because of the presence of the operator $D_0^\me$, though.

Let us return to the question of the implementation within the
framework of feed-forward networks.\\
We assume that the input layer knows to compute
                     $r_0(z)=  D_0\xi (z) - f_0(z) $ from
the input $\xi $. This is easily implemented within the standard
feed-forward framework by adding an extra layer.
Initial fine grid relaxation sweeps to smoothen $r_0$ are also
easily implemented.
The nodes $x$ of the hidden layer will receive input
$r_1(z)= \sum_z C(x,z)r_0(z) $ that is computed from the smoothened
version of $r_0$.

Let us imagine that we do not use the exact solution $e_1 $ to the
coarse grid equation~\eq{8.5},
                      but content ourselves with an approximation
$\Xi (x)$ to $-e_1(x)$ which comes from a single  Jacobi
relaxation sweep. Then the node $x$ is asked to compute the output
\be \Xi (x)= - d_1(x)^{-1}\ r_1(x) \es , \ee
from its input $r_1(x)$,
which it can legally do. $d_1$ is the diagonal part of $D_1$, viz.
$$
  D_1f(x)=d_1(x)f(x) +
               \mbox{ contributions }
  \propto f(y) \mbox{ with }          y\not=x\es .
$$
We imagine that there are additional connections
(not shown in figure
\eq{NEURAL_TWOGRID}  )
                                                 of strength $1$
 from input nodes to output nodes which represent the same lattice point
$z$. The total input to the output node $z$ is then
\be O(z) =\xi (z) + \sum_x\A (z,x)\ \Xi (x) \approx
    \xi (z) - \sum_x\A (z,x)\ e_1(x) \ , \ee
and this is also the output from the output node $z$.

Updating of connection strengths $C(x,z)$ and $\A (x,z)$ proportional
to the gradient of the cost functional $E$ leads to a standard
\it back-propagation learning rule \rm as described in
\cite{Palmer}. As is discussed there (Sect. 6.1)
                                the same network - or rather a
bidirectional version of it - can be used to compute the necessary
adjustments.

There is still one nonstandard feature.
       The network should also learn
$d_1(x)^{-1}$. This becomes a standard learning problem (to be solved
by the back-propagation learning rule)                  if we think
of duplicating the hidden layer, connecting
the duplicates of site $x$ by "neurons"
of connection strength $d_1(x)^{-1}$.

We discussed a twogrid for simplicity. Generalization to a multigrid is
obvious and will be presented elsewhere \cite{NEURAL_NEW}.
\section{Blocked Gauge Fields}
 Remember that the gauge covariant Laplace operator $\Lap$ for gauge
 field $U$ has kernel ($a = 1$)
 \begin{equation}
   \Lap (z,w) =  \left \{ \begin{array}{ll}
    -2 d     & \quad \mbox{ if $z=w$}, \\
    U(z,w)   & \quad \mbox{ if $z$\,n.n.\,$w$}
    \enspace  .         \end{array}  \right.
 \label{Lapkernel}
 \end{equation}
in d dimensions.
 And similarly for the staggered Dirac operator%
 \footnote{We use the conventions $\eta_{-\mu} = - \eta_{\mu}$
           and $e_{-\mu} = - e_{\mu}$. $z$ is called odd if
the sum of its integer coordinates is odd.}
 \begin{equation}
   \Dirac (z,w) =  \left \{ \begin{array}{ll}
    \eta_{\mu} (z)\,U(z,w)   & \quad \mbox{ if $w=z + \eh e_{\mu}$,
                                  $\mu = -d , \ldots , d, \mu\neq 0$},\\
    0        & \quad \mbox{ otherwise}
    \enspace  .         \end{array}  \right.
 \label{Dirackernel}
 \end{equation}

 A blockspin definition for matter fields gives us the effective
 operators on coarser layers, $\Lap_{\subs{eff}} = \A^* \Lap \A$ and
 $\Dirac_{\subs{eff}} = \A^* \Dirac \A$.
 Given $\Lap_{\subs{eff}}$ or $\Dirac_{\subs{eff}}$\,, we have
 candidates for gauge fields $\UB$ on the block lattice,
 \begin{eqnarray}
  & \UB(x,y)  = & \Lap_{\subs{eff}}(x,y) \quad \mbox{for $x,y$
    nearest neighbours}
   \nonumber\\
  \mbox{resp.\ } &
  \UB(x,y) =  & \Dirac_{\subs{eff}}(x,y)\,/\,\eta_{\mu}(x)
                \quad \mbox{if $y=x + L_b \eh e_{\mu}$} \es .
 \label{Ublock}
 \end{eqnarray}
 These expressions for
              $\UB(x,y)$ have the correct gauge covariance property.

 We add some comments on the Dirac case.
 Since we use odd $L_b$, the staggered block lattice can be regarded
 as a sublattice of $\Lambda$ in which the pseudoflavor of a site does
 not depend on whether we regard it as a site of the fundamental or
 the block lattice, and the same is true of its property of being even
 or odd.
                                         The kernel $\A $ will always
 have the property that $\A (z,x)=0$ unless $z$ and $x$ are both even
 or both odd. Therefore $\Dirac_{eff}$ connects even to odd sites,
 The factor $\eta_{\mu}(x)^{-1}$ can be regarded as defined by
 regarding $x$ as a point in a sublattice of the fundamental lattice.
 This agrees with the natural definition on the block lattice. Note also
 that   $\eta_{\mu}(z)$ is independent of the $\mu$-component
 $z_{\mu}$) of $z$.

 The blocked gauge fields $\UB(x,y)$ are actually ``dielectric'' gauge
 fields~\cite{MacDLGT}.
 They may be brought back into the gauge group by multiplying with
 a positive real number (for $U(1)$ and $SU(2)$)
 or with a positive matrix (for $U(N), N> 2$).

 A different choice of blockspin for gauge fields was introduced
 by Balaban and used in rigorous
 renormalization group work~\cite{GAUGE_BALABAN}.
\section{Simulation Methods for the Computation of Effective Actions}
The standard Monte Carlo simulation techniques for the computation
of functional integrals can be used to compute effective actions, and
in particular effective potentials.

There are three methods to take into account the $\delta $-function
constraints.

1.)
{\bf Gaussian blockspins.} Use a Gaussian in place of the
     $\delta$-function in the constraint. This definition of a block
spin transformation is very convenient for numerical work. But it
                    has some inherent dangers which have been
discussed in the literature (cp.~\cite{PINN}):
                                                    Fictitious
marginal operators may appear and cloud the renormalization group flow.
Some numerical results are found in~\cite{PINN}.

2.)
{\bf Simulations at fixed blockspin $\Phi$}~\cite{PINN}.
                                                   Do the updating
so that $\Phi$ is fixed. Compute
 $$     \langle\partial \H/\partial\Phi(x)\rangle_\Phi $$
This determines $\H_{eff}(\Phi)$ up to a constant.

For linear blockspin transformations, the quantity
  $    \partial \H/\partial \Phi(x)$ is defined as follows:
\be
   \frac{\partial \H     }{\partial \Phi (x)}= \int_z
   \frac{\partial \H (\varphi )}{\partial\vp (z)}\A (z,x)
\ee
    so that the action $\H $ changes by
    $ \delta \H =
   \frac{\partial \H}{\partial\Phi (x)}\delta \Phi (x)$
when $\vp $ is changed by an amount
     $ \delta \vp(z)=\A(z,x)\delta\Phi(x)$.
Simulations with fixed block spin are simulations for an auxiliary
theory with an infrared cutoff $a_0^{-1}$. Therefore
the critical slowing down will be governed by the ratio
             $a_0/a$  which is usually appreciably less than the
correlation length in units of $a$.

This method is general, but it is costly when one wants to compute
more than effective potentials.

3.)
{\bf Method of ``fluctuating coupling
constants''}~\cite{EVALUATE_CARGESE,EVALUATE_PHI4}
This method is applicable in cases such as Higgs models with and without
gauge fields, where the
effective action depends only on matter fields
$\Phi$ which are related by a linear blockspin definition to matter
fields which appear \it polynomially \rm in $\H$.\\

Given $\vp , U$ then the high frequency field $\zeta$ is defined.
A sequence of configurations  $(\vp_a , U_a)$ yields therefore
a sequence $ (\zeta_a,U_a),\ a=1,2,\dots,N_{conf} $. Expanding
\be
 \H(U_a,\zeta_a+\A\Phi) = \sum_m g_a^m \Phi^m
\ee
defines a sequence
 $g_{a}^m$ of "fluctuating coupling constants". They are stored. At
the end of the run, the effective potential can be recovered from them.
For a 1-component field
\be e^{-\H_{eff}(\Phi )}= N_{conf}^{-1}\sum_a Z_a^{-1}\exp
\left[ - \sum_m g_a^{m} \Phi^m \right] \ee
where $Z_a$ is the 1-dimensional integral
\be    Z_a = \int d\Phi \exp
\left[ - \sum_m g_a^{m} \Phi^m \right] \ee
One may choose to expand around another point than $\Phi =0$ instead.

In principle, fluctuation coupling constants could also be used to
encode the information about full effective actions. But in this case
$g_a^m$ would be  functions of $m$ arguments $x_1,\dots , x_m $ on the
block lattice. Only few of them could possibly be stored.
\subsubsection*{Multigrid Updating of Matter Fields}
The multigrid method provides
 a tool to fight critical slowing down
in Monte Carlo simulations.
    It uses or is motivated by constructs that we discussed. Therefore
we would like to mention it at least briefly.

Imagine that the blockspin $\Phi^j$ attached to some block lattice
$\L^j$ of lattice spacing $a_j$ is changed at site $x\in\L^j$ by
$\delta\Phi^j(x)$. This will induce a change of the matter field
$$
\delta\vp(z) = \A^j(z,x)\delta\Phi^j(x) \es .
$$
$\A^j$ should be thought of as minimizing the kinetic energy subject
to some constraint (such as $C\vp=\Phi$),
                    therefore it should be {\it smooth in $z$}
(no jumps). If we do not specify the averaging kernel $C$, there are no
further restrictions on $\A $ .

     Similar updatings are possible for fields subject to constraints
such as  unit vectors in $\sigma$-models, spins in $CP^n$-models
etc. For instance
\be
     \vp (z) \mapsto \exp\left[ \A(z,x)\ \epsilon\right]\ \vp (z)
\label{UnigridUp}\es ,
\ee
where the matrix $\epsilon $ is an element of the Lie algebra of the
symmetry of the model.

In the ``unigrid'' approach of Hasenbusch,
Mack and Meyer~\cite{EVALUATE_O3}, hypercubes $x$ of different side
lengths are
considered and updatings of the form~\eq{UnigridUp} are performed
with smooth kernels $\A(z,x)$ which vanish for $z$ outside $x$.
The positions of the hypercubes $x$ are chosen at random in the lattice.
It turns out to be important for the effectiveness of the method
that the selected hypercubes $x$ of one length have some overlap.

The unigrid method proved
very successful for 2-dimensional asymptotically free models:
                        The critical exponent $z$ which governs
critical slowing down is nearly zero in this case.
References are~\cite{EVALUATE_CP3}
for the CP$^3$-model and~\cite{EVALUATE_O3} for the
 $O(3)-$model.
It is essential for the success that these models have no mass
parameter~\cite{GrabPinn}.
The effective potential for 4-dimensional $\varphi^4-$theory was
computed using a similar updating scheme~\cite{EVALUATE_PHI4},
but with disjoint
         hypercubes $x$ which form a lattice, and with kernels $\A $
which had support on hypercubes $x$ and on their nearest neighbours.
\subsection*{Acknowledgements}
Support by Deutsche Forschungsgemeinschaft is gratefully
acknowledged.
  For providing resources and advice,
we are indebted to HLRZ/J\"ulich and its staff.
  G.M. and M.S. thank the organizers of the 31.IUKT in Schladming
for their efforts in creating the pleasant and stimulating
atmosphere during the school.
One of us (M.S.) thanks Thorsten Ohl
for gnuidance.
\section*{Appendix: Con\-straint
          Ef\-fec\-tive Po\-ten\-tial with Gauge Fields
in Fi\-nite Vol\-ume}
 \addcontentsline{toc}{section}{\protect\numberline{}
                 {\hspace*{-18pt}
                  Appendix: Con\-straint Ef\-fec\-tive Po\-ten\-tial
                  with Gauge Fields in Fi\-nite Vol\-ume}}
The result of the 1-loop calculation of Sect.~\ref{EFFECTIVE}
reads as follows (for details see \cite{PALMA_3})
\be
 U_{V} (\Phi) = U^{(0)}_{V} (\Phi) %
 + \frac{1}{2 V}\,\Tr\,\left[\ln \left\{ \frac{1}{2}\,\frac{\delta}{%
\delta \Arzmu}\,\frac{\delta}{\delta A_{r \bar{z}\nu}}\left(S^{(q)}_{g}%
[\Phi, A] \right)\right\}\right]_{A = 0}
 \label{veff-vol}
 \ee
with
\be
 U_{V}^{(0)} (\Phi) =  U_{cl} (\Phi) + \frac{1}{2 V} {\sum_{n}}'\ln%
 \left(\hat{k}^{2}_{n} + U''_{cl} (\Phi)\right) + \frac{3}{2 V}%
 {\sum_{n}}'\ln (\hat{k}^{2}_{n} + U'_{cl} (\Phi) /\, \Phi)
\ee
\be
 S^{(q)}_{g} [\Phi, A]  = S^{q}_{g} [A] +
                          U^{(2)}_{V}\, [\Phi, A]
\ee
where
\be
 \begin{array}{ll}
 U^{(2)}_{V} (\Phi, A) = &
   \frac{1}{4 V} (U'_{cl} (\Phi)/ \Phi)
 {\ds_{n}}' \frac{1}{(\Kn)^{2}} {\ds_{r}} \Arno
 + \frac{1}{4 V} {\ds_{n}}' (\Gamma
 _{n\phi_{0}}(\Phi) + 3 \Gamma_{n\phi_{r}}(\Phi))\\
 \\
  & {\ds_{r}}(A^{2}_{r n n} + \frac{1}{\Kn} \Arno)
 - \frac
{1}{4 V^{2}}{\ds_{n,m}}'(2 \Gnp \Gmp + \Gamma_{n \phi_{0}}(\Phi)\Gmp\\
 \\
 &+\Gnp \Gamma_{m \phi_{0}}(\Phi) ) {\ds_{r}} | A_{r n m} |^{2}%
   + \frac{\lambda_{0}}{12}  \Phi^{2} {\ds_{n}}'%
(4 \Gnp + \Gamma_{n \phi_{0}}(\Phi))\\
\\
&\frac{1}{(\Kn)^{2}} {\ds_{r}}\Arno%
\end{array}
\ee
 with
\be  \left\{
\begin{array}{l}
 U_{cl} (\Phi) = \frac{1}{2}\, m^{2}_{0}\,%
\Phi^{2} + \frac{\lambda_{0}}{4!} \Phi^{4},\mbox{\ \ \ }\Kn = 4%
\sum_{\mu} \sin^{2} (\frac{\pi}{L} n_{\mu}),\ \ \frac{L}{2} < n_{\mu}%
\leq \frac{L}{2} \\
\\
\Gamma%
 _{n \vp_{0}}(\Phi) = (\Kn + U''_{cl} (\Phi))^{-1},\mbox{\ \ \ \ \ }%
\Gamma%
 _{n \varphi_{r}}(\Phi) = ( \Kn + U'_{cl} (\Phi) / \Phi%
 )^{-1}
\end{array}  \right.
\ee
 $\Phi$ the absolute value of the covariant magnetization
 of~\eq{cov_magne} and $\ %
 A_{r n o}, A^{2}_{r n n}, A_{r n m}$ defined by
\be
  \left\{ \begin{array}{ll}
  (A^{2}_{r})_{oo} & = \frac{1}{V}\,{\ds_{z \mu}} \psi^{*}_{0}\, (z)%
  \Arz \Arz \psi_{0} (z - \mu) \\
  \\
  (A^{2}_{r})_{no} & = \frac{1}{V} {\ds_{z \mu}} \psi^{*}_{n}\, (z)%
  \Arz \Arz \psi_{0} (z - \mu) \\
  \\
  A_{r n m} & = \frac{1}{V}\,{\ds_{z \mu}}\,\psi^{*}_{n}\,(z) \Arz\,%
  \psi_{m}\,(z - \mu)\\
  \\
  \Arno & = A_{r n o} A^{*}_{r n o}
  \end{array} \right.
\label{21.3}
\ee
Here,
 $\psi_{n} (z) = \exp (\frac{2 \varphi i}{L} nz)$
 and the eigenvalue $\epsilon_{0}$ is
 given by~\eq{g2}. $U'_{cl}$ and $U''_{cl}$ are the first and
 second derivative of $U_{cl}$, and prime $'$ on the sum means that
 the $n = 0$ term is omitted.

 The trace in~\eq{veff-vol}
 involves summation over the discrete $O(4)$-%
 matrix indices and a sum over space-time points on a lattice
 $\Lambda$.

\end{document}